\documentclass[12pt]{JHEP3}

\usepackage{amssymb, amsmath, amsopn, amsthm}

\usepackage{epsfig}

\usepackage{graphicx}

\def\makeatletter{\catcode`\@=11}
\makeatletter
\def\mathbox#1{\hbox{$\m@th#1$}}%
\def\math@ccstyles#1#2#3#4#5#6#7{{\leavevmode
      \setbox0\mathbox{#6#7}%
      \setbox2\mathbox{#4#5}%
      \dimen@ #3%
      \baselineskip\z@\lineskiplimit#1\lineskip\z@
      \vbox{\ialign{##\crcr
             \hfil \kern #2\box2 \hfil\crcr
             \noalign{\kern\dimen@}%
             \hfil\box0\hfil\crcr}}}}
\def\mathaccstyles{\math@ccstyles\maxdimen}
\def\maththroughstyles{\math@ccstyles{-\maxdimen}}
\def\unity%
 {\maththroughstyles{.45\ht0}\z@\displaystyle {\mathchar"006C}\displaystyle 1}

\title{ Holography of Dyonic Dilaton Black Branes}

\author{Kevin Goldstein$^1$, Norihiro Iizuka$^2$, Shamit Kachru$^3$, Shiroman Prakash$^{4}$, ~\\ ~\hspace{-.595cm}{\normalsize \bfseries \sffamily Sandip P. Trivedi$^{4}$ and Alexander Westphal$^{3}$}

 ~\\
 $^1$National Institute for Theoretical Physics (NITHeP), \\
 School of Physics and Centre for Theoretical Physics, \\
 University of the Witwatersrand, WITS 2050, Johannesburg, South Africa \\

$^2$Theory Division, CERN, CH-1211 Geneva 23, Switzerland \\

$^3$Department of Physics and SLAC\\
Stanford University, Palo Alto, CA 94305 \\
 
$^4$Tata Institute for Fundamental Research  \\
Mumbai 400005, India\\

}

\abstract{ We study black branes carrying both electric and magnetic charges in Einstein-Maxwell theory coupled to a dilaton-axion in asymptotically
anti de Sitter space.   After reviewing and extending earlier results for the case of electrically charged branes, we characterise the thermodynamics of
magnetically charged branes.  We then focus on dyonic branes in theories which enjoy an $SL(2,R)$ electric-magnetic duality.    Using $SL(2,R)$, we are
able to generate solutions with arbitrary charges starting with the electrically charged solution, and also calculate
 transport coefficients.   These solutions all exhibit a Lifshitz-like near-horizon geometry.  The system behaves as expected for a charged fluid
in a magnetic field, with  non-vanishing Hall conductance and vanishing DC longitudinal conductivity at low temperatures.
 It's response is characterised by  a cyclotron resonance at a 
frequency proportional to the magnetic field, for small magnetic fields. 
Interestingly, the DC Hall conductance is related to the attractor value of the axion.
We also study the attractor flows of the dilaton-axion, both in cases with and without an additional modular-invariant scalar potential.  
The flows exhibit
intricate behaviour related to the duality symmetry. 
Finally,  we   briefly discuss  attractor flows in 
more general dilaton-axion theories which do not enjoy $SL(2,R)$ symmetry.}

\preprint{
{\normalsize CERN-PH-TH-2010-152, SU-ITP-10/22} \\
{\normalsize TIFR/TH/10-18, WITS-CTP-55} \\
}

\def\be{\begin{equation}}

\def\ee{\end{equation}}

\def\bea{\begin{eqnarray}}

\def\eea{\end{eqnarray}}

\def\tw{r}

\begin{document}

\tableofcontents

\section{Introduction}

The AdS/CFT correspondence provides us with a marvellous new tool for the study of strongly coupled field theories. 
There is hope and  excitement that these developments  might lead to a better understanding of  some 
 quantum critical theories occurring in Nature, for example    
 in superfluid-insulator transitions or  in cuprate materials which exhibit  high $T_c$ superconductivity \cite{Sachdev, Hartnoll, Herzog, McGreevy, Gary}.
Strong  repulsion due to charge  is believed to   play an important role in some of these critical theories.  
Modelling such repulsion on the gravity side leads one to consider extremal black brane gravitational solutions  whose mass
essentially arises entirely from  electrostatic  repulsion.  
In fact  extremal black branes/holes are fascinating objects in their own right, and have been at the centre of much of the progress
in understanding black holes in string theory. 
A possible tie-in with experimentally accessible  quantum critical phenomena only adds to their allure. 

With these general motivations in mind, charged dilatonic black branes in AdS space-times were discussed  in \cite{GKPT}. 
Earlier work on the subject had mostly dealt with the case of the Reissner-Nordstrom black brane. This is   interesting in many ways 
 but suffers, in the context of our present  motivations,  from  one  unpleasant feature. An extremal 
Reissner-Nordstrom black hole, which is the zero temperature limit of this system,  has a large entropy.  This feature seems quite unphysical,
and in the non-supersymmetric case it is almost  certainly a consequence of the large $N$ limit in which the gravity description  is valid.
It leads one to the worry that  perhaps other properties, for example transport properties like conductivity etc., calculated using 
 this brane would also receive large corrections away from the large $N$ limit, leaving the Reissner-Nordstrom system to be of only limited
interest in the present context.\footnote{It has recently been suggested that perhaps the large entropy of the Reissner-Nordstrom brane can be
interpreted as arising from some analogue of a ``fractionalized Fermi liquid" phase in the boundary theory \cite{Subir}.  Some support for the existence of such
a phase, at least in some AdS/CFT dual pairs, accrues from explicit lattice models with localised fermions in string constructions, where $AdS_2$ regions
arise from bulk geometrization of the lattice spins \cite{KKY}.  While this is an intriguing possibility, here we adopt the view that it would be good
to find natural models without the large ground-state entropy.   
Another, complementary approach to the entropy problem is developed in \cite{d'Hoker}.}

In the dilatonic case, in contrast, it was  found that the extremal electrically charged brane has zero entropy \cite{GKPT}.
Its near-horizon geometry shows that the dual theory in the infra-red has scaling behaviour of Lifshitz  type \cite{KLM} with a non-trivial 
dynamical exponent $1/\beta$ (where $\beta < 1$ is determined by the details of the dilaton coupling to the gauge field), and with additional logarithmic violations. Departures from extremality give rise to an entropy density $s$
 growing as a power law $s\sim T^{2\beta}$, with a positive specific heat.   The optical conductivity, for small   frequency compared to the 
chemical potential $\mu$, is of the form $Re(\sigma) \sim \omega^2$, with  the power law dependence being independent of the dynamical exponent $\beta$. 

In this paper, we  continue the study  of  extreme and near-extreme dilatonic black branes. We find that in the electric case
at small frequency and temperature, when  $ \omega \ll T\ll \mu$, the conductivity  is $Re(\sigma) \sim T^2$ 
(with  an additional delta function at $\omega=0$). The field theory we are studying has a global Abelian symmetry and the conductivity determines
the transport of this global charge.   
To characterise the field theory better it is useful to 
gauge this  global symmetry, then turn on a background magnetic field and study the resulting response. 
This also corresponds to turning on a magnetic field in the gravity dual.
 
Once we are considering a bulk magnetic 
field it is also natural to add an axion in the bulk theory.\footnote{It is reasonable to believe that varying the boundary value of the axion corresponds
to adjusting the value of a Chern-Simons coupling in the boundary theory \cite{Kraus}; we briefly expand on this comment in \S6.}
A particularly interesting case is when the bulk theory has an $SL(2,R)$ symmetry. \footnote{This symmetry is expected to only be approximate
and would receive corrections beyond the classical gravity approximation; for instance, in many quantum string theories, it is broken to $SL(2,Z)$ non-perturbatively.} In this case the behaviour of a system carrying both electric and magnetic charges
can be obtained from the purely electric case using an  $SL(2,R)$ transformation. One finds that the system 
is diamagnetic.  Under an $SL(2,R)$ transformation the dilaton-axion,
$\lambda= a+ i e^{-2\phi}$, transforms like $\lambda \rightarrow {\tilde{a} \lambda + b \over c\lambda + d}$. 
It  turns out that the   two complex combinations of the  conductivity  $\sigma_{\pm}=\sigma_{yx}\pm i  \sigma_{xx}$ also\footnote{
The conductivities $\sigma_{xx}, \sigma_{yx}$, are frequency dependent and complex thus $\sigma_{\pm}$ are not complex conjugates of each other.}
   transform in the same way,
$\sigma_{\pm}\rightarrow {\tilde{a}\sigma_{\pm}+b\over c \sigma_{\pm}+d}$, allowing us to easily determine them. 
An important check is that the  resulting Hall conductivity at   
zero frequency is $\sigma_{yx}={n\over B}$ where $n,B$ are the charge density and magnetic field, and the 
longitudinal conductivity at zero 
frequency vanishes. These results   follow simply   from Lorentz invariance in the presence of a magnetic field.  
An interesting feature of  our results is that the  DC Hall conductivity agrees with the attractor value of the axion.
This is in accord  with  expectations  that 
 the axion  determines  the coefficient 
of the Chern-Simons coupling in the boundary theory, which in turn determines the Hall conductivity. 

Besides the electric  conductivity  we also calculate the thermoelectric and thermal conductivity for a general system carrying
both electric and magnetic charges. These are related to the electric conductivity by Weidemann-Franz type relations which are quite analogous to those obtained in the non-dilatonic case \cite{HKMS, HH}.
As was noted above,  
 the electric conductivity behaves quite similarly as a function of temperature or frequency 
 in the dilaton-axion  and non-dilatonic  cases. The Weidemann-Franz type relations
then lead to the   thermoelectric and thermal conductivities  also  behaving in a similar way in these cases. 

We also  discuss the attractor flows for the axion and dilaton in these dyonic branes, and find intricate flow diagrams whose properties are governed by
the SL(2,R) symmetry.  In cases with a suitable SL(2,Z) invariant potential, we find that for fixed charges there can be multiple attractor points, governing
different basins of attraction in field space.

Finally we 
 consider a more general class of bulk theories containing a dilaton-axion but without $SL(2,R)$ symmetry.
For some range of parameters we find that the deep infra-red geometry is an attractor and changing the asymptotic value of the
 axion does not lead to a change in this geometry. Outside this parametric  range, however, 
the attractor behaviour appears to be lost and a small change in the asymptotic value of the axion results in a solution which becomes 
increasingly different in the infrared.



This paper is structured as follows. \S2 contains a review of the salient points in \cite{GKPT}. \S3 contains a discussion of the DC conductivity at finite temperature in the purely electric case. \S4 contains a discussion of the case with only a magnetic field 
and no charge. This is a warm up for the more general discussion with  both electric and magnetic charges which is   analysed for a system with 
$SL(2,R)$ invariance in  \S5. 
Additional discussion of conductivity and other transport coefficients in this case is contained in \S6. 
Attractor flows in these systems, both with and without a bare potential for the dilaton-axion, are discussed in \S7.
Some more general systems without 
$SL(2,R)$ symmetry are discussed in \S8. 
Finally \S9 contains some concluding comments. 
Supporting material appears in the appendices.

\section{Review of earlier results}

Here we summarise some of the results of \cite{GKPT}.
Consider a four-dimensional system consisting of a  dilaton coupled to a gauge field and gravity with action
\begin{equation}
\label{ouraction}
S = \int d^4x \sqrt{-g} \left(R - 2(\nabla \phi)^2 - e^{2\alpha\phi} F^2 - 2\Lambda\right)~.
\end{equation}
$\Lambda =-{3 \over L^2}$ is the cosmological constant.  We  will often set  $L=1$ in the discussion below.

The metric of a black brane has the form
\begin{equation}
\label{ansatz}
ds^2 = -a(r)^2 ~dt^2 + a(r)^{-2} ~dr^2 + b(r)^2 ~(dx^2 + dy^2)
\end{equation}
For an electrically charged brane the gauge field is
\begin{equation}
\label{gaugefield}
e^{2\alpha \phi} F = {Q\over b(r)^2} ~ dt \wedge dr.
\end{equation}

The extremal black brane is asymptotically $AdS_4$ and  characterised by two parameters, the charge $Q$ and $\phi_0$ - the 
asymptotically constant  value of the dilaton.
In the extremal case, the near-horizon region is universal and independent of both these parameters, due to 
the attractor mechanism \footnote{The curvature scale in the near-horizon region is set by the cosmological constant
 $\Lambda$.}. The metric is of the Lifshitz form \cite{KLM}\footnote{See also \cite{Koroteev}.}
\begin{equation}
\label{nhmetricl}
ds^2=-(C_2 r)^2 dt^2+ {dr^2\over (C_2 r)^2}+r^{2\beta}(dx^2+dy^2),
\end{equation}
with dynamical exponent
\begin{equation}
\label{dynamicexponent}
z={1\over \beta}.
\end{equation}
The near-horizon solution is valid when
\be
\label{condnh}
r \ll \mu 
\ee
where $\mu \propto \sqrt{Q}$ is the chemical potential. 

The dilaton in the near-horizon region  is
\be
\label{dilaton}
\phi=-K {\rm log}(\tw)~.
\ee
The constants which appear in the metric and  dilaton above   are given in terms of $\alpha$, 
the coefficient in the dilaton coupling eq.(\ref{ouraction}):
\be
\label{constants}
C_2^2 =  {6\over(\beta+1)(2\beta+1)}~, \beta = {({\alpha \over 2})^2 \over {1 + ({\alpha \over 2})^2}}~,
~K = {{\alpha \over 2} \over {1 + {({\alpha \over 2})}^2}}.
\ee
This class of solutions, but with different asymptotics than those of interest to us, was discussed in \cite{Taylor} (the solutions
there were asymptotically Lifshitz, and have strong coupling at infinity; for other asymptotically Lifshitz black hole solutions, see
\cite{LifBH, Takayanagi}).

The entropy of the extremal black brane vanishes. For a near-extremal black brane  the temperature 
dependence of entropy and other thermodynamic quantities is essentially determined by 
the near-horizon region.  (For a careful discussion of how the global embedding affects the thermodynamics,
see appendix A of \cite{Perlmutter}; see also the recent paper \cite{Amanda} for a recent discussion of how the non-extremal branes embed into AdS.).

The bulk theory above is dual to a $2+1$ dimensional boundary theory which is a  CFT with a globally conserved $U(1)$ symmetry. The electrically charged black brane is dual to the boundary theory in a state with
constant charge density determined by $Q$.

 The black brane geometry can be used to calculate transport coefficients in the boundary theory.
In particular, the  real part of the longitudinal electric conductivity  ($Re(\sigma)\equiv \sigma_{xx}= \sigma_{yy}$)
at zero temperature and small frequency is found to be
\footnote{There is a delta function Drude peak at $\omega=0$ in addition which we have subtracted.}
\be
\label{cond1}
Re(\sigma)=C {\omega^2\over \mu^2}.
\ee
Here $C$ is a constant which depends on $\alpha$ and $\phi_0$.
We note that the frequency dependence of $Re(\sigma)$ is universal and is independent of $\alpha$.  
The conductivity is dimensionless in $2+1$ dimensions. This fixes the dependence on $\mu$ - the chemical potential-  once the dependence on $\omega$ is known.

More generally, at finite temperature and frequency, $\sigma$ is a function of two dimensionless variables
 $\sigma({T\over \mu}, {\omega \over \mu})$.
Eq.(\ref{cond1}) gives the leading dependence when $T \ll \omega \ll \mu$. 
We also note that in the purely electric case the Hall conductivity $\sigma_{xy}$ vanishes. 

\section{The DC conductivity}
In this section we calculate the leading behaviour of the conductivity, $\sigma$, 
 when 
\be
\label{pararange}
\omega  \ll T \ll \mu.
\ee
Our analysis will closely follow the discussion in \S3 of \cite{GKPT} (which itself used heavily the results of \cite{Roberts}). 
We  consider a perturbation in $A_x$, which mixes with the metric component $g_{xt}$, impose 
in-going boundary conditions at the horizon,  and then carry out a matched asymptotic expansion which  determines
 the behaviour near the boundary and hence the conductivity. 
We skip some of the details here and emphasise only the central points.\footnote{The result of this section has also been obtained in \cite{Kiritsis}, which 
appeared while our paper was
being readied for publication.  Other related papers which appeared recently include \cite{Cadoni, list}.}

The leading behaviour of the conductivity in the parametric range eq.(\ref{pararange})
will turn out to be 
\be
\label{conddc}
Re(\sigma)=C' {T^2\over \mu^2}
\ee
This is independent of $\omega$. The DC conductivity defined 
as the limit $\omega \rightarrow 0$ of the above formula then just gives eq.(\ref{conddc}) as the result. 
Actually there is an additional delta function contribution at $\omega=0$; we will comment on this more in the following subsection. 
 $C'$ in eq.(\ref{conddc}) is a constant that depends on $\phi_0$. 
 
We begin with a coordinate system in which the metric is, 
\be
\label{metricb}
ds^2=-ge^{-\chi}dt^2+{d\tilde{r}^2\over g}+\tilde{r}^2(dx^2+dy^2)
\ee
and  define a variable $z$
\be
\label{defz}
{\partial \over \partial z} = e^{-\chi/2}g {\partial \over \partial \tilde{r}}.
\ee
One can then show that the variable 
\be
\label{defpsi}
\Psi=f(\phi) A_x
\ee
satisfies  a Schr\"odinger  equation, 
\be
\label{sepsi}
-{\Psi}''+V(z) \Psi=\omega^2 \Psi, 
\ee
where prime indicates derivative with respect to $z$, 
and $f^2(\phi) = 4 \, e^{2 \alpha \phi}$, as discussed in eq.(3.10) of \cite{GKPT}. 
The potential is
\be
\label{defpot}
V(z)={f''\over f}+ g^{-1} f^2 e^\chi (A'_t)^2
\ee

Comparing the $g_{tt}, g_{rr}$ components in eq.(\ref{ansatz}), eq.(\ref{metricb}) we see that 
\be
\label{rela}
g e^{-\chi}=a^2, \ \ {dr \over d{\tilde r}}=e^{-\chi/2}
\ee
so that  
\be
\label{zr}
{\partial \over \partial z}= a^2 {\partial \over \partial r}.
\ee
The potential eq(\ref{defpot}) is
\be
\label{fformpot}
V={f''\over f}+{a^2Q^2\over b^4 f^2}.
\ee

In the near-boundary region,  $\Psi$ takes the form
\be
\label{nboundary}
\Psi=(D_1+D_2) + i\omega(-D_1+D_2) z.
\ee
The resulting flux is 
\be
\label{resflux}
{\cal F}\sim |D_1+D_2|^2\omega Re(\sigma).
\ee

We are interested here in a slightly non-extremal black brane. 
This has a near-horizon metric 
\be
\label{nhmetric}
ds^2=-C_2^2 r^2(1-({r_h\over r})^{2\beta+1}) dt^2 + {dr^2\over C_2^2 r^2(1-({r_h\over r})^{2\beta+1})}+ r^{2\beta}(dx^2+dy^2),
\ee
The temperature is
\be
\label{tempne}
T \sim r_h.
\ee 
The dilaton is the same as in the extremal case. 
The near-horizon form of the metric above   is valid  for $r \ll \mu$. 
The temperature dependence  of the conductivity 
is essentially determined by the near-horizon region, as long as ${T\over \mu}\ll 1$.
This is similarly to what happens for the frequency dependence when ${\omega\over \mu} \ll 1$. 

In  the near-horizon region $r_h$ is the only scale, as we can see from eq.(\ref{nhmetric}).  
It is therefore convenient, in the discussion below, to rescale variables by appropriate powers of  $r_h$. 
We define
\be
\label{hw}
\hat{r}={{r}\over r_h} 
\ee
\be
\label{hata}
\hat{a}^2\equiv {a^2\over r_h^2}= C_2^2\hat{r}^2  (1- ({1\over \hat{r}} )^{2\beta+1} )
\ee
and
\be
\label{hatz}
{\partial \over \partial \hat{z}}\equiv {1\over r_h} {\partial \over \partial  {z}}= \hat{a}^2 {\partial \over \partial \hat{r}}
\ee

The Schr\"odinger equation then becomes,
\be
\label{seb}
-{d^2\Psi \over d\hat{z}^2} + \hat{V} \Psi = {\omega^2\over r_h^2} \Psi
\ee
where the rescaled potential, $\hat{V}$, is dependent on the rescaled variable $\hat{z}$ alone  without any 
additional dependence on $r_h$ . 

Very close to the horizon, $\hat{V}$ goes to zero and we have
\be
\label{vchorizon}
\psi \sim e^{\bigl(-i\omega(t+z)\bigr)} = e^{-i\omega t} e^{-i ({\omega\over r_h} \hat{z})}
\ee
resulting in the flux
\be
\label{fluxnh}
\cal{F}\sim \omega.
\ee
From eq.(\ref{resflux}), eq.(\ref{fluxnh}) we see that the conductivity is
\be
\label{condd}
Re(\sigma) \sim {1\over |D_1+D_2|^2}.
\ee

Now, consider the region of the near-horizon geometry where 
\be
\label{twocond}
{\mu\over T} \gg \hat{r} \gg 1 ~. 
\ee
Since the temperature is small eq.(\ref{pararange}), these conditions are compatible. 
 In this region  the temperature dependent terms in the metric are subdominant and $a^2\simeq C_2^2 r^2$. Eq.(\ref{hatz})  
then leads to 
\be
\label{valhatz}
\hat{z}=-{1\over C_2^2 \hat{r}}
\ee
and eq.(\ref{fformpot}) to a potential,
\be
\label{potform}
\hat{V}={c\over \hat{z}^2},
\ee
with the constant
\be
\label{valc}
c=2.
\ee

Now since the frequency is even smaller than the temperature, eq.(\ref{pararange}), $\omega/T \ll 1$ and eq.(\ref{twocond}) and eq.(\ref{tempne})
  imply that 
\be
\label{secondcond}
\hat{r} \gg {\omega \over r_h}.
\ee
In terms of $z$ this becomes 
\be
\label{twocondz}
 {1\over \hat{z}^2}\gg ({\omega\over r_h})^2.
\ee

It follows that the
  the frequency  term in the Schr\"odinger equation eq.(\ref{seb}) is subdominant
compared to the potential term in this region.  
The resulting solution becomes  
\be
\label{soln}
\Psi\simeq \hat{z}^{1/2}({a_1 \over \hat{z}^\nu}+b_1\hat{z}^\nu)
\ee
with
\be
\label{valnu}
\nu=\sqrt{c+{1\over 4}}.
\ee
From the condition $\hat{r} \gg 1$ and eq.(\ref{valhatz}) we see that in this region 
\be
\label{onemore}
|\hat{z}| \ll 1.
\ee
As a result,  
the first term on the rhs of eq.(\ref{soln}) dominates \footnote{This would not be true if $a_1$ was suppressed compared to $b_1$ by a 
power of $\omega$. However, this does not happen, as we discuss further in Appendix A.}
  giving
\be
\label{condpsi2}
\Psi\sim a_1 (r_h z)^{({1\over 2}-\nu)}
\ee
Here we have used the fact that $\hat{z} = r_h z $. 

We have seen above that once $r$ lies in the region which  meets the condition eq.(\ref{twocond}) both 
the temperature and frequency effects can
 be neglected. Moving outwards towards the boundary this continues to be true  all the way till the 
near boundary region. This region is  described in Step 1 of
\S3.2.2\ in \cite{GKPT}. As a result, one gets 
\be
\label{vald}
D_1\sim D_2\sim r_h^{({1\over 2}-\nu)}
\ee
From eq.(\ref{condd}), eq.(\ref{tempne}), eq.(\ref{valc}) and eq.(\ref{valnu}),  this gives
\be
\label{condda}
Re(\sigma)\sim (r_h)^{(2\nu-1)}\sim T^{2\nu-1}\sim T^2.
\ee
The dependence on $\mu$ then follows from dimensional analysis, leading to eq.(\ref{conddc}).

Finally we note that it is simple to see that 
 the Hall conductivity continues to vanish at finite temperature as well. 

\subsection{The pole in $Im(\sigma)$ and related delta function in $Re(\sigma)$ }
The real part of $\sigma$ has a delta function contribution at $\omega=0$, which arises because the system has a net charge and 
it is transported in a   momentum conserving manner. 
A Kramers-Kronig relation  relates the delta function  to a pole in  the imaginary part of $\sigma$. 
It will be important to keep track of this pole and the related delta function when we turn to the discussion of the system in a magnetic field,
 so let us discuss it  in some more detail here. 
We will rely on the analysis in \S3 of \cite{GKPT}.

As discussed in \S3.1 of \cite{GKPT}, following \cite{Roberts}, the conductivity is given in terms of the reflection coefficient ${\cal R}$ by
\be
\label{condref}
\sigma={1-{\cal R}\over 1+{\cal R}}
\ee
(the extra term in eq.(3.12) of \cite{GKPT} drops out since $f'(0)$ vanishes like $z^3$ towards the boundary).

Now in the notation of \S3.2 of \cite{GKPT}  close to the boundary $\Psi$ is
\be
\label{defpsib}
\Psi= D_1 e^{-i\omega (t +  z)} +D_2 e^{-i\omega (t -z)}, 
\ee
giving
\be
\label{condreff}
\sigma={D_1-D_2\over D_1+D_2}.
\ee

The coefficients  $D_1, D_2$ can be related to  $E_1,E_2$ which govern the solution in 
the not-so near boundary region. This region is defined in Step 1 of \S3.2.2 in \cite{GKPT}
and corresponds to taking $|\omega|\ll z \ll 1$. The coefficients $E_1,E_2$ are defined in     eq.(3.30) of 
\cite{GKPT}, by
\be
\label{e1e2}
D_1+D_2=E_1, \ \ D_1-D_2=i {E_2\over \omega},
\ee
giving from eq.(\ref{condreff})
\be
\label{valcondref}
\sigma=i {E_2\over E_1} {1\over \omega}.
\ee

Now $E_2,E_1$ are obtained by starting from the near horizon region where in-going boundary conditions are imposed and integrating out towards the boundary.  
The Schr\"odinger equation is real. And 
in the zero temperature case discussed in \cite{GKPT}, the solution to leading order
in   the near horizon region is given in the equation after equation (3.32) there. We see that  is of the form,  
$\psi = C z^{1/2-\nu}$. Integrating this out towards the boundary will give $E_2/E_1$ to be real and of order unity in units of the
 chemical potential.  
Similarly at non-zero temperature in the parametric range eq.(\ref{pararange}) the solution in the near -horizon region eq.(\ref{twocond})
 is given by eq.(\ref{condpsi2}). 
Once again integrating out towards the boundary gives $E_2/E_1$ to be real and of order unity. 
Thus we learn that near $\omega=0$ 
\be
\label{imsigmac}
Im(\sigma) = C'' {\mu \over \omega}
\ee
where $C$ is a coefficient of order unity and we have restored the $\mu$ dependence on dimensional grounds. 
As a result there is indeed a pole at $\omega=0$ in $Im(\sigma)$, and hence as discussed above a delta function in $Re(\sigma)$ at $\omega=0$. 

In the presence of disorder the frequency dependence changes,  
${1\over \omega} \rightarrow {1\over \omega+ i/\tau_{imp}}$ \cite{HKMS}, and the pole acquires an imaginary part. 
As a result the delta function peak in $Re(\sigma)$ is broadened out as will be discussed further in \S6.

\section{Purely magnetic case}

Next, as a warm-up for general dyonic branes, we consider the case of a black brane which carries only magnetic charge. The dyonic case, with a bulk axion as well, 
will be investigated in subsequent sections. 
The action is given by eq.(\ref{ouraction}), but now we are interested in the case where the gauge field strength is 
\be
\label{magneticf}
F=Q_m dx\wedge dy.
\ee
It is easy to see that the equations of motion for the system are invariant under a duality transformation
which keeps the metric invariant \footnote{This is the Einstein frame metric.} and  takes
\be
\label{dta}
\phi \rightarrow -\phi, \ \ F_{\mu\nu}\rightarrow e^{2\alpha \phi} \tilde{F}_{\mu\nu}.
\ee
Here 
\be
\label{deftilde}
\tilde{F}_{\mu\nu}={1\over 2} \epsilon_{\mu\nu\rho\sigma} F^{\rho\sigma}.
\ee
So we see that starting from the electric case eq.(\ref{gaugefield}), we get to the magnetic one eq.(\ref{magneticf})
after the duality transformation discussed above.
The value of $Q_m$ is
\be
\label{magdualch}
Q_m=Q.
\ee

As a result, the metric for the extremal  magnetic case 
in the near horizon region   is still of the Lifshitz form eq.(\ref{nhmetricl}). 
To avoid confusion we denote the dilaton after duality by $\phi'$ in the subsequent discussion; it is given by
\be
\label{dilmag}
\phi'=K \log(r)
\ee
where the constants which appear in the metric and in the dilaton continue to be given by eq.(\ref{constants}).
The gauge coupling is $(g')^2 = e^{-2\alpha \phi'}$. From eq.(\ref{dilmag}), eq.(\ref{constants}) we see that 
the theory  now gets driven to strong coupling, $(g')^2 \rightarrow \infty$,
 near the horizon, and if a string embedding is possible
this would mean that quantum loop effects would get important near the horizon. 
By considering a slightly non-extremal black brane such effects can be controlled. 

The behaviour of the dilaton can also be understood in terms of the effective potential \cite{attractors}. 
In general, with electric and magnetic charges  the effective potential is (from eq.(2.19) of \cite{GKPT}):
\be
\label{effpot}
V_{eff}=e^{-2\alpha \phi}Q_e^2+e^{2\alpha \phi} Q_m^2
\ee 
Since after duality,  $Q_e=0, Q_m=Q$, we get,
\be
\label{efmag}
V_{eff}=Q^2 e^{2\alpha \phi'}
\ee
so that the minimum does indeed lie at $e^{2 \alpha \phi'} \rightarrow 0$, or equivalently 
$e^{-2\alpha \phi'}\rightarrow \infty$. 

In mapping the magnetic case to the boundary theory it is best to think of weakly gauging the global U(1) symmetry 
of the boundary theory.  Then the magnetic case corresponds to turning on a constant magnetic field in the 
boundary theory. The electric-magnetic duality therefore has an  interesting consequence. In the electric case, 
the electric  field is a normalisable mode and  corresponds to a state in  the boundary theory  
at constant number  density or chemical potential. In contrast, in the magnetic case, the magnetic field  
 is a non-normalisable mode and  corresponds to changing the Lagrangian of the 
 boundary theory.   

The metric in the slightly non-extremal case is also unchanged by duality and hence given in the near-horizon region 
by eq.(\ref{nhmetric}). 
We now elaborate on the  resulting thermodynamics.

\subsection{Thermodynamics}
Let us begin by briefly reviewing the purely electric case. 
From the Maxwell term in the action
\be
\label{emact}
S_{em}=-\int d^4x \sqrt{-g}e^{2\alpha \phi} F_{\mu\nu}F^{\mu\nu}
\ee
using standard techniques in AdS/CFT and the definition of $Q$, eq.(\ref{gaugefield}), 
 we learn that the  the charge density  $n$ 
in the boundary theory   is 
\be
\label{stech}
n=4 Q.
\ee

A purely electric system satisfies the thermodynamic relation
\be
\label{pet}
TdS=dE+pdV-\mu dN  
\ee
From this relation, using electric-magnetic duality, one can obtain the thermodynamic quantities in the magnetic case. 
For this purpose it is convenient 
to take the independent thermodynamic variables in the electric case to be 
$(E,V, T,n )$, since these can be mapped directly to the independent variables $(E,V,T,Q_m)$ in the magnetic case. 
Here $Q_m$ is the magnetic field \footnote{The   magnetic field is usually denoted by $H$ or $B$, but $Q_m$ is more 
natural for us in view of the duality transformation.}.
Since the Einstein frame action is duality invariant $(E,V,T)$ are  left
 unchanged in going from the electric to the magnetic case.
And  from eq.(\ref{stech})
and  eq.(\ref{magdualch}) it follows that $n \rightarrow 4 Q_m$. Thus,
 the four independent variables can be easily mapped to one another.

Expressing the number $N=nV=4QV = 4 Q_m V$ we get from eq.(\ref{pet}) in the electric case that 
\be
\label{dt}
TdS=dE + (p-4 \mu Q) dV -  4 \mu V dQ_m.
\ee
 Comparing eq.(\ref{dt}) with the standard  thermodynamic relation in the purely magnetic case (as discussed in e.g. Reif, ${\it Fundamentals~ of~ Statistical
 ~and ~Thermal~ Physics,}$ 11.1.7)
\be
\label{timag}
TdS=dE+pdV + M dH~,
\ee
and noting that the magnetic field is $Q_m$ in our notation, 
we get  that the magnetisation is 
\be
\label{mag}
M=-4 \mu V
\ee
and the pressure in the magnetic case is
\be
\label{pmag}
p_{mag}= p_{el}-4 \mu Q = p_{el}+ {MH\over V}.
\ee
In the electric case the chemical potential is a function of the energy density $\rho, T, n$, $\mu(\rho,T,n)$. In the 
formulae above for the magnetic case, eq.(\ref{mag}), eq.(\ref{pmag}),  the chemical
 potential should now be interpreted as a function of $\rho, T, Q_m$ given by $\mu(\rho,T, 4Q_m)$. 
 
It is worth discussing  the extremal situation in the magnetic case further.  
The energy density (see eq.(2.52) of  \cite{GKPT}) is given by
\be
\label{rhoext}
\rho=C Q^{3/2} e^{-3\alpha \phi_0/2} =  C (V_{eff 0 })^{3/4}
\ee
where we have used the definition of the effective potential in eq.(\ref{effpot}).
The subscript ``$0$'' on $V_{eff}$ indicates that it is to be evaluated at $\infty$, where the dilaton takes value $\phi_0$.

The chemical potential is 
\be
\label{valmu}
\mu={\partial \rho \over \partial n}={3\over 8} C Q^{1/2} e^{-3\alpha \phi_0 / 2} = {3\over 8} C (Q_m)^{1/2} e^{3\alpha \phi'_0 / 2}
\ee
where we have used eq.(\ref{magdualch}) and eq.(\ref{dta}). 
We see from eq.(\ref{mag})
that the magnetisation is opposite to the magnetic field.  As a result, the susceptibility for this system is 
negative, and the  theory is  diamagnetic. 

 Using $p_{el}=\rho/2$, (\cite{GKPT} eq.(2.53)), the  pressure in the magnetic case is
\be
\label{pmagthree}
p_{mag}=-\rho=-C H^{3/2} e^{3\alpha \phi'_0 / 2}
\ee
It seems puzzling  at first that that this is negative, since one would expect the boundary theory to be 
stable. This turns out to be   a familiar situation 
 in  magnetohydrodynamics, see the discussion around eq.(3.10) in \cite{HKMS}. In the presence of a magnetic field the pressure and 
spatial components of stress energy are different and related by
\be
\label{pmagb}
T^{xx}=T^{yy}= p_{mag}-{M H \over V}.
\ee
Stability really depends on  the sign of $T^{xx}$, which determines the force acting on the system. 
From eq.(\ref{pmag}),  we see that $T^{xx}=p_{el}$, and is thus positive. 
\footnote{In fact this had to be true since $T^{xx}, T^{yy}$ are  duality invariant and in the electric case $p_{el}=T^{xx}=
T^{yy}$.}

\subsection{Controlling the flow to strong coupling}

 We saw above, eq.(\ref{efmag}), that for the magnetic case $e^{2\alpha \phi'}\rightarrow 0$ and thus 
the gauge coupling $g^2= e^{-2\alpha \phi'}$
gets driven to strong coupling at the horizon.  In a string theory embedding one would expect the string coupling to become large and thus quantum corrections to become important near the 
horizon. To control these corrections one can consider turning on a small temperature and dealing with  the near-extremal brane instead. 
From eq.(\ref{magdualch}), eq.(\ref{tempne}), and eq.(\ref{constants})   we see  that if the temperature is $T\sim r_h$ the coupling
 at the horizon is 
\be
\label{dilmagh}
e^{-2\alpha \phi'}\sim {1\over T^{4 \beta}}
\ee
The only other dimensionful quantity in the boundary theory is the magnetic field, so the dependence on magnetic field can be fixed by dimensional analysis.  An explicit
 bulk analysis also shows that this dependence is correct.  
In addition there is a dependence on the asymptotic value of the dilaton $\phi_0'$. It is easy to see that  $\phi_0'$ only enters in 
the combination
$Q_m e^{\alpha \phi_0'}$ with the magnetic field and as  $(\phi'-\phi_0')$ with the varying dilaton. 
This is enough to fixed the $\phi_0'$ dependence of eq.(\ref{dilmagh}) and we get
\be
\label{tdimh}
e^{-2\alpha \phi'} \sim e^{-2\alpha \phi_0'} ({Q_m e^{\alpha \phi_0'} \over T^2})^{2\beta}.
\ee
For the temperature to be small and the brane to be near-extremal, 
\be
\label{condnhaa2}
T^2 \ll Q_m e^{\alpha \phi_0'}.
\ee
Thus to make $e^{-2\alpha \phi'} \ll 1$ we need to adjust the asymptotic value of dilaton and start with a theory which is at very weak coupling
\be
\label{condnhaa3}
e^{-2\alpha \phi_0'}\ll ({T^2 \over Q_m e^{\alpha \phi_0'}})^{2\beta}.
\ee
Once this is done the coupling will continue to be small all the way to the horizon. 

\subsection{Dyonic case with only dilaton}
Most of this section has dealt with the purely magnetic case. Below we will turn to a dyonic system with an axion. 
Before doing so though let us briefly discuss the dyonic case in the presence of only a dilaton without an  axion. 
From eq.(\ref{effpot}) we see that the dilaton now has the attractor value $\phi_*$ with,
\be
\label{dnd}
e^{2\alpha\phi_*}=|{Q_e\over Q_m}|.
\ee
From the equations of motion it then follows that the metric component $b^2$, eq.(\ref{ansatz}), at the horizon is 
\be
\label{ddb}
b_h^2 \sim \sqrt{V_{eff}(\phi_*)} \sim \sqrt{|Q_e Q_m|}.
\ee
The resulting entropy is then 
\be
\label{resendd}
s \propto b_h^2/G_N \sim C \sqrt{|Q_e Q_m|}
\ee
where $C \sim L^2/G_N$ is the central charge of the $AdS_4$.
As has been discussed above the purely electric case has no  ground state degeneracy. 
 Once a magnetic field is also  turned on we see that such a degeneracy does arise.  By itself this is not surprising.
However, the  resulting entropy formula, eq.(\ref{resendd}), is quite intriguing and understanding it better  should provide important 
clues for  the microscopic dual of the purely  dilatonic case.

\section{The $SL(2,R)$ invariant case}
 In this section we discuss a theory which has $SL(2,R)$ duality symmetry, in the presence of an axion,  with action \footnote{
In our conventions $\tilde{F}^{\mu\nu}={1\over 2} \epsilon^{\mu\nu\rho\kappa}F_{\rho\kappa}$ and $\epsilon^{\mu\nu\rho\sigma}$ has 
a factor of  ${1\over \sqrt{-g}}$ in its definition, thereby making the axionic coupling independent of the metric. We have chosen conventions $\epsilon_{trxy} > 0$.}, 
\be
\label{acsl}
S=\int d^4x \sqrt{-g}[R-2\Lambda-2(\partial \phi)^2-{1\over 2} e^{4\phi} (\partial a)^2 -e^{-2\phi}F^2 - a F \tilde{F}].
\ee
Comparing with eq.(\ref{ouraction}) we see that the gauge coupling function here corresponds to taking $\alpha=-1$. 
We will mostly follow the notation of \cite{DDD} below (see also \cite{Sen})  and denote the complexified dilaton-axion by 
\be
\label{cdil}
\lambda=\lambda_1+i\lambda_2 = a + i e^{-2\phi}.
\ee

It is easy to see that under an $SL(2,R) $ transformation 
\be
\label{matsl2r}
M=\begin{pmatrix}\tilde{a}& b \cr c & d\cr
\end{pmatrix}
\ee
 which takes
\be
\label{transgauge}
F_{\mu\nu}  \rightarrow   F_{\mu\nu}'  =  (c \lambda_1+d) F_{\mu\nu}-c \lambda_2 \tilde{F}_{\mu\nu}
\ee
and
\be
\label{transdil}
\lambda  \rightarrow   \lambda'={\tilde{a} \lambda + b \over c\lambda + d} \\
\ee
while keeping the metric invariant, 
the equations of motion are left unchanged. 
(This is discussed for example in \cite{Sen} around eq.(18) with $(ML)_{ab}\rightarrow -1$).  
Note that we are denoting $M_{11}=\tilde{a}$  and the axion by $\lambda_1\equiv a$ to avoid confusion.  Also, since $M$ is an element of $SL(2,R)$
\be
\label{detrel}
\tilde{a}d-bc=1.
\ee
 
Thus starting from the purely electric case where only the dilaton is non-trivial and carrying out a general 
duality transformation, we can obtain
a dyonic brane with both axion and dilaton excited.
In the discussion below we will follow the conventions established above of referring to  parameters obtained after duality with 
a prime superscript. 

The starting electric brane is characterised by four parameters:  a mass $M$, a charge $Q$, and asymptotic values of the dilaton and axion,
 $\lambda_{20} \equiv e^{-2\phi_0}$, $\lambda_{10} \equiv a_0$. The axion is radially constant.  
The $SL(2,R)$ transformation adds three additional parameters, \footnote{$det(M)=1$ so there is 
one constraint among the $4$ matrix elements.}
resulting in a $7$ parameter set of solutions. Two of these parameters are redundant, though, since  
the general dyonic brane solution only has only  $5$ independent parameters:
$M', Q'_e, Q'_m, \lambda'_{20}, \lambda'_{10}$. 
This redundancy can be removed by  setting $\lambda_{10}=0$ in the electric case, and also setting $Q=1$
\footnote{More correctly the scaling symmetry allows one to set $|Q|=1$.}.
In the discussion below we will set $\lambda_{10}=0$,  but not necessarily set $Q=1$. 

The gauge field can be written in terms of the electric and magnetic charges as follows
\be
\label{gfd}
F'={(Q_e'-Q_m'\lambda_1') \over b(r)^2}(\lambda_2')^{-1} dt\wedge dr+ Q_m' dx\wedge dy
\ee
It can be seen that $Q_e', Q_m'$ being constant solves the gauge field equations of motion and Bianchi identities. 
From eq.(\ref{gfd}) we see that 
\be
\label{gftc}
F'_{xy}=Q_m'.
\ee
Using eq.(\ref{transgauge}) this gives, 
\be
\label{magp2}
Q_m'
= -c \lambda_2 \tilde{F}_{xy}=c Q.
\ee
Similarly from eq.(\ref{gfd}) we see that 
\be
\label{eqgf}
\lambda_2' F'_{tr}= {(Q_e'-\lambda_1'Q_m') \over b(r)^2}.
\ee
And eq.(\ref{transgauge}) now gives  
\be
\label{gftb}
F'_{tr}=(c\lambda_1+d) F_{tr}-c \lambda_2 \tilde{F}_{tr}=d F_{tr}=d {Q\over \lambda_2 b(r)^2}
\ee
where we have used eq.(\ref{magp2}) and the fact that $\lambda_1=a=0$ and $\tilde{F}_{tr}=0$ in the electric case. 
 Together these imply
\be
\label{valqep}
Q_e'=({\lambda_2'\over \lambda_2} d + \lambda_1' c) Q.
\ee
Using eq.(\ref{transdil}), and relation $\tilde{a} d -bc=1$ then gives
\be
\label{fqep}
Q_e'=\tilde{a} Q.
\ee

It is now easy to see that the effective potential, which is given by
\be
\label{effpotgen}
V_{eff}'=(Q_e'-Q_m'\lambda_1')^2(\lambda_2')^{-1}+ (Q_m')^2 \lambda_2' ~,
\ee
is in fact duality invariant and  thus equal to its value in the purely electric frame,   
\be
\label{eptwo}
V_{eff}={Q^2 \over \lambda_2}. 
\ee

Thermodynamic quantities of  a system carrying  electric charge in a magnetic  field satisfy the relation 
\be
\label{tem}
TdS=dE+pdV-\mu dN + M  d Q_m
\ee
We will be particularly interested in the extremal case where the $TdS$ term vanishes. 
Writing $E=\rho V, N=n V$ we get in this case,
\be
\label{exthermo}
(d\rho-\mu dn + {M\over V} dQ_m) V + (\rho-\mu n + p) dV=0
\ee
From this it follows that both, 
\be
\label{reltha}
(d\rho-\mu dn + {M\over V} dQ_m)=0
\ee
and 
\be
\label{relthb}
(\rho-\mu n + p)=0.
\ee

We are interested in applying these relations  to the dyonic case obtained after duality. 
The energy density is duality invariant, since it can be extracted from the Einstein frame metric which is duality invariant. 
Thus we get, 
\be
\label{endd}
\rho' = \rho=C(V_{eff 0})^{3/4}=C [(Q_e'-Q_m'\lambda_{10}')^2(\lambda_{2 0}')^{-1}+ (Q_m')^2 \lambda_{20}']^{3/4} 
\ee
The subscript ``$0$'' on $V_{eff}$  and the moduli indicates that the effective potential must be evaluated at $\infty$
where the moduli take values $\lambda_{20}'\equiv e^{-2\phi_0'}, \lambda_{10}'\equiv a_0'$. 
Straightforward manipulations then give us that
\be
\label{newmu}
\mu'={1\over 4}{\partial \rho'\over \partial Q_e'} = {3 C \over 8} (V_{eff 0})^{-1/4}({Q_e'-\lambda_{10}'Q_m'\over \lambda_{20}'})
\ee
where we have used the fact that $n'=4 Q_e'$.
The magnetisation per unit volume is
\be
\label{muv}
{M' \over V}= -{\partial \rho' \over \partial Q_m'}= -{3 C \over 2 (V_{eff 0 })^{1/4} \lambda_{20}'}
[Q_m'(\lambda_{20}'^2+\lambda_{10}'^2)-\lambda_{10}'Q_e']
\ee
and  the pressure is 
\be
\label{newp}
p'=\mu' n' - \rho'=-{C\over (V_{eff 0})^{1/4} \lambda_{20}'}[(Q_m')^2 (\lambda_{20}'^2+\lambda_{10}'^2) 
-\frac{1}{2} ({{Q_e'^2 + \lambda_{10}' Q_e'Q_m' }})]
\ee
In eq.(\ref{newmu})-(\ref{newp}) the moduli take their values at infinity.  From eq.(\ref{muv}) it follows  that 
the susceptibility is negative, and thus the system is diamagnetic.
 From eq.(\ref{newp}) we see that the pressure can   be positive or negative. 
The stress energy tensor  component $T^{xx}=T^{yy}= \rho/2$ and is always positive. 

Finally, we discuss the compressibility of this system. This is defined to be
\be
\label{defk}
\kappa=-{1\over V} {\partial V \over \partial p}|_{T Q_m N}
\ee
The partial derivative on the rhs is to be evaluated at constant temperature $T$, magnetic field $Q_m$ and total number
$N=V n$. For a system of fermions which has precisely enough particles to fill an integer number of Landau levels, reducing the volume while
keeping the magnetic field $Q_m$ fixed would change the available number of states in the occupied Landau levels.  But
since the total number of fermions is not being changed in the process, and there is a large gap to the next available Landau level,
this cannot happen without significant energetic cost, and as a result the compressibility vanishes. This happens for example in quantum Hall systems.
For our case,  from eq.(\ref{reltha}) eq.(\ref{relthb}) we have that
\be
\label{evalcom}
{\partial p \over \partial V}|_{T Q_m N} = n {\partial \mu \over \partial V}|_{T Q_m N} =  n
{\partial \mu \over \partial n}|_{T Q_m } ({\partial n \over \partial V})_N.
\ee
This gives
\be
\label{fkappa}
\kappa={1\over n^2} ({\partial n \over \partial \mu})|_{T Q_m}.
\ee
From the expression for $\mu'$ eq.(\ref{newmu}) it is easy to see that
$({\partial \mu' \over \partial n'})|_{T Q'_m}$ cannot   go to infinity for finite $V_{eff}$, and non-vanishing 
$\lambda_{20}$ and  thus the compressibility
cannot vanish except in extreme limits.
So the system at hand cannot  become incompressible, except when $V_{eff} \to 0$ and/or $e^{-2\phi} \to 0$.  We will see
that some of the natural attractor flows in $SL(2,R)$ invariant theories do result in incompressible states of holographic matter. 
 
\section{Conductivity in the $SL(2,R)$ invariant case}
We now turn to calculating the conductivity in the $SL(2,R)$ invariant case discussed in the previous section. 
The conductivity is defined as follows
\begin{eqnarray}
j_x& = & \sigma_{xx} F_{tx}+ \sigma_{xy} F_{ty} \label{conddef} \\
j_y & = & \sigma_{yx} F_{tx} + \sigma_{yy} F_{ty} \label{conddef2}.
\end{eqnarray}
Under a rotation by $\pi/2$, which is a symmetry of the system,  $(x,y)\rightarrow (y,-x)$. Transforming all quantities appropriately in the above equations we learn that  
\be
\label{relsigma}
\sigma_{xx}=\sigma_{yy}, \ \ \sigma_{xy}=-\sigma_{yx}.
\ee
Thus there are two independent components in the conductivity tensor. 
In the discussion below we will use the notation
\be
\label{defs12}
\sigma_1={\sigma_{yx} \over 4}, \ \ \sigma_2={\sigma_{xx}\over 4}.
\ee
 
Below we will use the bulk description to calculate $j_x, j_y$, in terms of the boundary value of gauge fields.
From the resulting equations we will find that the two  complex combinations   
\be
\label{defs}
\sigma_+=\sigma_1 + i \sigma_2
\ee
\be
\label{defsbar}
\sigma_-= \sigma_1-i\sigma_2
\ee
both 
transform in the same way as the axion dilaton under an $SL(2,R)$ transformation. Namely 
\be
\label{tsl}
\sigma_{\pm} \rightarrow {\tilde{a} \sigma_{\pm} + b \over c \sigma_{\pm} + d}
\ee
under the transformation eq.(\ref{matsl2r}). 
Note that the conductivity  components $\sigma_{xx}, \sigma_{yx}$ are in general complex. Thus $\sigma_+ $ and $\sigma_-$ are not complex 
conjugates of each other. Starting from the purely electric case, for which the  conductivity has already been obtained above, 
and using the  transformation properties, eq.(\ref{tsl}),  we can then easily obtain the conductivity 
for a general dyonic case.

The electromagnetic part of the bulk action is
\be
\label{emactb}
S_{em}=\int d^4x\sqrt{-g}[\lambda_2 F_{\mu\nu}F^{\mu\nu}-\lambda_1 F\tilde{F}].
\ee
In the subsequent discussion it is useful to work in a coordinate system where the metric takes the form
\be
\label{metrcond}
ds^2=a^2(-dt^2+dz^2)+b^2(dx^2+dy^2)
\ee
Asymptotically, the metric approaches $AdS_4$ and $a^2=b^2=z^{-2}$.
In the boundary theory, the current $\langle j_x \rangle$ can be obtained by
\be
\label{excu}
\langle j_x \rangle ={\delta \log(Z) \over \delta A_x}
\ee
The standard AdS/CFT dictionary then tells us that in the bulk,
\be
\label{bulkex}
\langle j_x \rangle = 4 [ \lambda_{2} F_{zx} -  \lambda_1 F_{ty}]_{z\rightarrow 0}
\ee
(here we have chosen conventions so that $\epsilon_{tzxy}>0$).
Similarly,
\be
\label{bjy}
\langle j_y \rangle =4[\lambda_2 F_{zy}+ \lambda_1 F_{ty}]_{z\rightarrow 0}.
\ee
In this section we will be mainly concerned with using these formula to calculate the conductivity.
For ease of notation in the subsequent discussion we will  not 
specify that the moduli and field strengths which appear are to evaluated at the boundary,
 $z\rightarrow 0$. 

From eq.(\ref{bulkex}), eq.(\ref{bjy}), eq.(\ref{conddef}), eq.(\ref{conddef2}) and eq.(\ref{defs}) we get
\begin{eqnarray}
\lambda_{2} F_{zx} -  \lambda_1 F_{ty}& = & \sigma_2 F_{tx}-\sigma_1 F_{ty} \label{condeq1} \\
\lambda_2 F_{zy}+\lambda_1 F_{tx} & = & \sigma_2 F_{ty}+\sigma_1 F_{tx}.  \label{condeq2}
\end{eqnarray}

A general $SL(2,R)$ transformation can be obtained by a product of two kinds of $SL(2,R)$ elements. 
The first, which we denote as $T_b$,  is of the form
\be
\label{tbdef}
\begin{pmatrix}  1 & b \cr 0 & 1 
\end{pmatrix}
\ee
And the second, which we denote by $S$, is
\be
\label{tsdef}
\begin{pmatrix} 0 & -1 \cr 1 & 0 
\end{pmatrix}
\ee
To show that  eq.(\ref{condeq1}), eq.(\ref{condeq2})
 transform in a covariant way under a general $SL(2,R)$ transformation, 
when $\sigma_{\pm}$ transform as given in eq.(\ref{tsl})  it is enough to show this for the transformations $T_b,S$. 
 
Under $T_b$ the field strength $F_{\mu\nu}$ does not change, eq.(\ref{transgauge}). The dilaton-axion transform as 
$\lambda_1\rightarrow \lambda_1 + b $, eq.(\ref{transdil}), and $\sigma_1 \rightarrow \sigma_1+b$, eq.(\ref{tsl}).
So we see that  eq.(\ref{condeq1}), eq.(\ref{condeq2}) are left unchanged. 
The lhs of  eq.(\ref{condeq1}) can be written as, 
\be
\label{translhs}
[\lambda_{2} F_{zx} -  \lambda_1 F_{ty}] = -{\lambda \over 2} (F_+)_{ty}-{\bar{\lambda}\over 2}(F_-)_{ty}
\ee
where $F_{\pm}=F\pm i \tilde{F}$. 
Under a general $SL(2R)$ transformation 
\begin{eqnarray}
\label{gslf}
F_+ & \rightarrow & F_+'=(c\lambda+d)F_+ \\
F_- & \rightarrow & F_-'=(c\bar{\lambda}+d)F_-. 
\end{eqnarray}
From this it follows that under $S$ the lhs of  eq.(\ref{condeq1}) goes to 
\be
\label{lhsfe}
[\lambda_{2} F_{zx} -  \lambda_1 F_{ty}]\rightarrow F_{ty}.
\ee

The RHS of eq.(\ref{condeq1}) can be written as 
\be
\label{rhscondeq1}
RHS=\sigma_2 F_{tx}-\sigma_1 F_{ty}= {1\over 2i}[\sigma_+(F_{tx}-iF_{ty}) -\sigma_-(F_{tx}+iF_{ty})].
\ee
Under a general $SL(2,R)$ transformation 
this becomes
\begin{eqnarray}
\label{genrhs}
RHS & \rightarrow &  {1\over 2i}[({ \tilde{a} \sigma_+ + b\over c\sigma_+ + d}) 
\{(c\lambda_1+d) (F_{tx}-iF_{ty}) - c\lambda_2(\tilde{F}_{tx}-i\tilde{F}_{ty})\}
 \nonumber \\
&& -({\tilde{a}\sigma_-+b \over c \sigma_-+d})\{ (c\lambda_1+d) (F_{tx}+i F_{ty})-c\lambda_2 (\tilde{F}_{tx}+i\tilde{F}_{ty}) \}]
\end{eqnarray}
From eq.(\ref{tsl}) after some algebra  it then   follows that under $S$  
\be
\label{rhstrans}
RHS \rightarrow {1\over \sigma_+ \sigma_-}[ \sigma_2(\lambda_1 F_{tx}+\lambda_2 F_{zy}) + \sigma_1 (\lambda_1 F_{ty}-\lambda_2 F_{zx})]
\ee
Using eq.(\ref{condeq1}), eq.(\ref{condeq2}) this becomes, 
\be
\label{gen2rhs}
RHS\rightarrow {1\over \sigma_+ \sigma_-}[\sigma_2(\sigma_1 F_{tx}+\sigma_2 F_{ty})+\sigma_1 (\sigma_1 F_{ty}-\sigma_2 F_{tx})] = F_{ty}
\ee
Thus the LHS and RHS of eq.(\ref{condeq1}) transform the same way if the  conductivity transforms as given in  eq.(\ref{tsl}).
A similar result can be obtained for eq.(\ref{condeq2}) thereby establishing that eq.(\ref{tsl}) is the correct transformation  law 
for $\sigma_{\pm}$.

Similarly, some algebra shows that if $\sigma$ transforms as in eq.(\ref{tsl}) the RHS of eq.(\ref{condeq1})
becomes, 
\be
\label{rhsfe}
\sigma_2 F_{tx}-\sigma_1 F_{ty} \rightarrow  {1\over \sigma_1^2+\sigma_2^2}[\sigma_2(\lambda_1 F_{tx}+\lambda_2 F_{zy})
-\sigma_1(\lambda_2 F_{zx}-\lambda_1 F_{ty})]
\ee
Upon using  eq.(\ref{condeq1}) this  gives 
\be
\label{rhsfet}
\sigma_2 F_{tx}-\sigma_1 F_{ty} \rightarrow F_{ty}
\ee
which is indeed  equal to the transformation of LHS, as seen in eq.(\ref{lhsfe}). 
Similarly  eq.(\ref{condeq2}) can also be shown to be covariant under $S$. 
This proves that  eq.(\ref{condeq1}), eq.(\ref{condeq2}) transform in a covariant manner under $SL(2,R)$. 

Since a general dyonic system can be obtained by starting from a purely electric one and carrying out an $SL(2,R)$ transformation,
we can now obtain the conductivity for the general dyonic case using eq.(\ref{tsl}). We will follow the conventions of the previous section and refer to quantities in the electric frame without a prime superscript and in the dyonic frame with a prime superscript. 
In the purely electric case we have $\sigma_{xy} = \sigma_{yx}=0$.
Thus $\sigma=i{\sigma_{xx} / 4}$. 
Also, it is enough to consider the case with the axion set to zero, $\lambda_1=0$,  in the electric frame. 
Thus $\lambda=i \lambda_2$. 
Then using eq.(\ref{tsl}) we get 
\be
\label{sigmaxx}
\sigma_{xx}'={\sigma_{xx}\over d^2+c^2({\sigma_{xx}\over 4})^2} 
\ee
and
\be
\label{sigmayx}
\sigma_{yx}'= 4 { \tilde{a} c ({\sigma_{xx}  \over 4})^2  + bd \over d^2+c^2({\sigma_{xx}\over 4})^2} \,.
\ee

To complete the analysis one would like to express the $SL(2,R)$ matrix elements which appear on the RHS  of eq.(\ref{sigmaxx}), 
eq.(\ref{sigmayx}) in terms of parameters in the dyonic frame.

As discussed in the previous section, the most general dyonic case can be obtained by starting with
a purely electric case with axion set to zero and $Q=1$. 
From eq.(\ref{fqep}), eq.(\ref{magp2}) we see that with $Q=1$
\be
\label{valac}
Q_e'=\tilde{a}, Q_m'=c.
\ee
The invariance of the effective potential gives, from eq.(\ref{effpotgen}), eq.(\ref{eptwo}),
\be
\label{l2}
\lambda_{20}^{-1}= (Q_e'-Q_m'\lambda_{10}')^2 (\lambda'_{20})^{-1}+(Q_m')^2 \lambda_{20}'.
\ee
This allows the asymptotic value of the dilaton in the electric frame to be expressed in terms of quantities in the dyonic frame. 
Using this and   eq.(\ref{valqep}) we learn that $d$ is 
\be
\label{valdb}
d={(Q_e'-\lambda_{10}'Q_m')\over (Q_e'-\lambda_{10}'Q_m')^2+(Q_m')^2(\lambda_{2 0}')^2} \,.
\ee
And then, finally, using the relation $\tilde{a}d-bc=1$ gives
\be
\label{valb}
b={\lambda_{10}'Q_e' - Q_m'(\lambda_{10}'^2+\lambda_{2 0}'^2)\over (Q_e'-\lambda_{10}'Q_m')^2+(Q_m')^2(\lambda_{2 0}')^2} \,.
\ee

\subsection{More on the conductivity }
The formulae obtained for the conductivity  eq.(\ref{sigmaxx}) eq.(\ref{sigmayx}) are valid in general. Let us  discuss the resulting  behaviour of the conductivity  at small frequencies and temperatures in 
the parametric range eq.(\ref{pararange}) more explicitly. 

To start it is useful to state the parametric range eq.(\ref{pararange}) in a duality invariant manner. 
The $SL(2,R)$ transformation with $b=c=0, \tilde{a}=1/d$ is a scaling transformation. 
Starting with the purely electric case, this SL(2,R) transformation yields $Q'_e=Q_e/d, Q_m'=0$. 
From eq.(\ref{newmu}), eq.(\ref{transdil}), it follows that the chemical potential and dilaton transform as
\be
\label{mudil}
\mu'=\mu d, \ \ \sqrt{\lambda_2'}=\sqrt{\lambda_2}/d,
\ee
 so that 
$\mu\sqrt{\lambda_2}$ is  invariant under the rescaling. This combination can in fact be expressed in terms of the  effective potential, which is duality invariant,  as
$\mu \sqrt{\lambda_2}\sim (V_{eff 0})^{1/4}$.  
The frequency $\omega$ and temperature $T$ are duality invariant. \footnote{The duality invariance of the temperature follows from that of the 
Einstein frame metric.} Thus  the duality invariant way to state the parametric range of interest is 
\be
\label{dipararange}
\omega \ll T \ll (V_{eff 0})^{1/4}.
\ee

In the purely electric case, the conductivity  to leading order is
\be
\label{locond}
\sigma_{xx}= C' {T^2\over \mu^2} + i C'' {\mu \over \omega}
\ee
Under the rescaling discussed in the previous paragraph, $\sigma_{xx}'=\sigma_{xx}/d^2$. From this and eq.(\ref{mudil}) it follows
that $C'$ is independent of $\phi_0$ while $C''\propto (\lambda_2)^{3/2}$.
Both $Re(\sigma_{xx})$ and $Im(\sigma_{xx})$ have corrections, which result in a fractional change of order $\omega^2$,   
\be
\label{corrsigma}
Re(\sigma_{xx})= C'{T^2\over \mu^2}(1+ O(\omega^2)), \ \  Im(\sigma_{xx}) = C''{\mu\over \omega} (1+ O(\omega^2)).
\ee
Plugging eq.(\ref{locond}) into the transformation laws eq.(\ref{sigmaxx}), eq.(\ref{sigmayx}),
 gives the conductivity for the general dyonic case. 

Let us consider the Hall conductance first. 
When the magnetic field is non-zero, $c\ne 0$ and the pole in the imaginary part of $\sigma_{xx}$ will dominate the low frequency 
behaviour. As a result, we get 
\be
\label{limhall}
\sigma'_{yx} = 4 {\tilde{a}\over c} + {\cal O}(\omega^2)
\ee
From eq.(\ref{valac}), eq.(\ref{stech}) we see that the leading behaviour is 
\be
\label{lhall}
\sigma'_{yx}= {n' \over Q_m'}
\ee
where $n', Q_m'$ are the charge density and the magnetic field respectively. 
This result in fact just follows from Lorentz invariance.

Intuitively, one would  expect that the DC value of the Hall conductivity  agrees with the coefficient of the Chern-Simons term 
of the dual field theory in the far infra-red, which in turn should be given by the value of the axion close to the horizon in the bulk. From (\ref{transdil})
 it follows that the axion
after the duality transformation is given by
\be
\label{dualaxion}
\lambda_1'={\tilde{a} c \lambda_2^2 + bd \over c^2 \lambda_2^2+d^2}
\ee
Near the horizon in the electric case $\lambda_2\rightarrow \infty$; thus, the attractor value of the axion is 
\be
\label{ataxion}
\lambda_{1*}'={\tilde{a} \over c}  
\ee
which is indeed proportional to the value of the Hall conductance  eq.(\ref{lhall}) (the factor of $4$, which is the proportionality
constant, follows from eq.(\ref{condeq1}), (\ref{condeq2})). 

Actually, it turns out that the ${\cal O}(\omega^2)$ terms in eq.(\ref{limhall}) can also be calculated reliably in terms of 
$C', C''$.  From eq.(\ref{sigmayx}) and eq.(\ref{locond}) we get that 
\be
\label{condayx}
\sigma'_{yx}={n' \over Q_m'}[1+ \omega^2\{-4({T^2 C'\over C''\mu^3})^2+{64 d \over \mu^2 n' (Q_m')^2 (C'')^2} \} 
+ {\cal O}(\omega^4)]
\ee

Next let us consider the longitudinal conductivity.
From eq.(\ref{sigmaxx}) we get,
\be
\label{condxx}
\sigma_{xx}'=-i {16\over (Q_m')^2}{\omega\over C''\mu}[1+ i {C'\over C''}{\omega T^2 \over \mu^3} + {\cal O}(\omega^2)]
\ee
Here $C',C''$ are the coefficients as given in eq.(\ref{locond}) and $\mu$ is the chemical potential in the electric theory.
We see that the longitudinal conductivity vanishes as $\omega\rightarrow 0$. This result also follows from Lorentz invariance
in the presence of a magnetic field. 
We also see that the imaginary part does not have a pole after the duality transformation; this shows  that there is no delta function at 
zero frequency in the real part of $\sigma_{xx}$. The absence of this delta function again is to be expected on general grounds, since
in the presence of the  background magnetic field, momentum is not conserved. 

It is worth comparing our results with the general discussion of conductivity for a relativistic plasma in \cite{HKMS}. 
From general reasoning based on linear response in magnetohydrodynamics it was argued in \cite{HKMS} (see also \cite{HH}) that at small frequency
\be
\label{sacsig}
\sigma_{xx}=\sigma_Q{\omega(\omega+i\gamma+i\omega_c^2/\gamma)\over [(\omega+i\gamma)^2-\omega_c^2]}
\ee
and
\be
\label{sacsxy}
\sigma_{xy}=-{n'\over Q_m'}({\gamma^2+\omega_c^2-2i\gamma\omega \over (\omega+i \gamma)^2-\omega_c^2})
\ee
Here $\sigma_Q, \gamma, \omega_c$ depend on the magnetic field $Q_m'$, $T$ and charge density $n'$.
$\gamma$ is the  damping frequency and $\omega_c$ is the cyclotron frequency. 
  Expanding in a power series for small $\omega$ gives
\be
\label{expsacsig}
\sigma_{xx}=-i{ \sigma_Q \omega  \over \gamma} [ 1+ {i \gamma \omega \over \gamma^2+\omega_c^2} + {\cal O}(\omega^2)]
\ee
and
\be
\label{expsxy}
\sigma_{xy}={n'\over Q_m'}[1+ {\omega^2\over \gamma^2 + 
 \omega_c^2}]
\ee
Comparing with eq.(\ref{condayx}), eq.(\ref{condxx}) we see \footnote{Our convention for $\sigma_{xy}$ differs from that of \cite{HH} by a sign.}
 that
\bea
\label{relussach}
{\gamma\over \gamma^2+\omega_c^2} & = & {C' T^2 \over C'' \mu^3} \cr
{1\over \gamma^2+\omega_c^2} & = & {64 d \over  n' Q_m'^2 C''} - 4({T^2 C'\over C''\mu^3})^2 \cr
{\sigma_Q\over \gamma} & = & {16 \over (Q_m')^2 C '' \mu}
\eea
These three relations determine $\sigma_Q, \gamma, \omega_c$ in terms of the parameters of our calculations. 
To express the answer in terms of the dyonic duality frame variables we should bear in mind that  
$d$ is given in terms of the charges etc in eq.(\ref{valdb}), $\mu \sqrt{\lambda_{20}}\sim (V_{eff0})^{1/4}$, and $\lambda_{20}$ is given in eq.(\ref{l2}).
Also while $C'$ is independent of $\lambda_{20}$, $C''\propto \lambda_{20}^{3/2}$. 


The equations in (\ref{relussach})  are  valid for small temperature eq.(\ref{dipararange}) and arbitrary $n', Q_m'$. 
It is easy to solve them and obtain $\sigma_Q, \gamma$ and $\omega_c$  in a small $T$ expansion. While we do not present the results in detail,
let us note that 
one finds at small  $T$ and  also small magnetic field $Q_m'$ that  
  $\sigma_Q, \gamma, \omega_c$ scale as,
\be
\label{scalingrel}
\sigma_Q \propto T^2, \ \ \gamma \propto (Q_m')^2 T^2,  \ \ \omega_c \propto Q_m'.
\ee
This qualitative behaviour is in agreement with the results of \cite{HH,HK} for the  
Reissner-Nordstrom black brane at small $\omega$ and $Q_m'$.  

\subsection{Thermal and thermoelectric conductivity}

There are two  transport coefficients related to the  conductivity, 
the thermoelectric coefficient $\alpha$ and the thermal conductivity $\kappa$.
Both should be thought of as tensors.   
These are defined by  the relations, 
\be
\begin{pmatrix}
\vec{J}\cr\vec{Q}
\end{pmatrix}=\begin{pmatrix} \bf{\sigma} & \bf{\alpha} \cr \bf{\alpha} T & \bf{\kappa} \end{pmatrix}
\begin{pmatrix} \vec{E} \cr -\vec{\nabla} T \end{pmatrix}
\ee
where $\vec{E}$ is the electric field, $\vec{\nabla}T$ is the gradient of the temperature, $\vec{J}$ is the electric current and 
$\vec{Q}$ is the heat current. 

It is easy to see, using the second law, that $Q^i$  is given by
\footnote{Ambiguities in the definition of the heat current can arise because entropy is not conserved. However they 
 enter in higher orders and are not important  in   linear response theory.}
\be
\label{dehc}
Q^i=T^{ti}-\mu J^i
\ee
where $T^{ti}$ is a component of the stress energy tensor and $\vec{J}$ is the electric current.\footnote{Some of the literature, e.g., \cite{HKMS}, 
defines transport coefficients in terms of currents where a magnetisation dependent term  is subtracted out. It is 
straightforward to relate
our answers to those obtained after such a subtraction.}

In AdS/CFT the source term corresponding to the electric field is a non-normalisable mode of the bulk gauge field $A_i$,
while the source corresponding to a thermal gradient $\nabla _i T$
 corresponds,  to  a combination of the non-normalisable mode for the metric component 
$g_{it}$ and $A_i$. By turning these on and calculating the response we can calculate the thermoelectric and thermal conductivities. 

\subsubsection{The thermoelectric conductivity}  
The thermoelectric coefficient 
$\bf{\alpha}$ can be determined by calculating the heat current  $\vec{Q}$ generated in response to an electric field
in the absence of a temperature gradient. In AdS/CFT we turn on a non-normalisable mode for $A_i$ and calculate the resulting value for 
$Q_i$. We will take the time dependence to be of the form $e^{-i\omega t}$ throughout.  
To begin we consider the $SL(2,R)$ case eq.(\ref{acsl}) but in fact our results will be quite general and we comment on this 
at the end of the subsection.

For a metric 
\be
\label{mettt}
ds^2=-a^2dt^2+{dr^2\over a^2} + b^2 (dx^2+dy^2) + 2 g_{xt} dx dt + 2 g_{yt} dy dt
\ee
and with action given by eq.(\ref{acsl}) 
we find that the $xt$ component of the trace-reversed Einstein equations gives
\be
\label{tree}
R_{xr} = 2 \lambda_2 (-F_{rt}F_{tx} g^{tt}+F_{ry}F_{xy} g^{yy} +  F_{rx} F_{xt}g^{xt}+F_{rt}F_{xt}g^{yt}
+F_{rt}F_{xy}g^{yt} )
\ee
with
\be
\label{valctx}
R_{xr}=-i\omega {\partial _r(g^{xx} g_{tx})\over 2 g_{tt} g^{xx}}.
\ee

The standard procedure to calculate the stress tensor in terms of the extrinsic curvature \cite{Skenderis, BK} gives
\be
\label{exttx}
T_{tx}=[ a \partial_r g_{tx} - 2 g_{tx} ] 
\ee
where the right hand side is to be evaluated close to the boundary as $r\rightarrow \infty$. 

While we  skip some of the steps in the analysis below, it is easy to see that 
close to the boundary, the leading behaviour on the rhs of eq.(\ref{tree}) comes from the first two terms.
Thus, we get close to the boundary from eq.(\ref{valctx}), eq.(\ref{tree})
\be
\label{newvalc}
-i\omega {\partial _r(g^{xx} g_{tx})\over 2 g_{tt} g^{xx}} \simeq 2 \lambda_2 (-F_{rt}F_{tx} g^{tt}+F_{ry}F_{xy} g^{yy})
\ee

Substituting  eq.(\ref{gfd}) for the field strength then yields,
\be
\label{expt}
T_{tx}={2\over i \omega}[-2{(Q_e'-\lambda_{10}'Q_m') \over a} E_x' + 2 \lambda_2' Q_m'F_{ry}' a ]
\ee
Some of the notation we have adopted here is potentially confusing.  The superscript prime here denotes 
a dyonic configuration with both electric and magnetic charge as in the previous
sections. In particular,  
the variable $\lambda_{10}'$ denotes the asymptotic axion in the system with both electric and magnetic charge. The variable   
 $a$ in the equation above stands for the redshift factor in the metric. 

Using the relation between the variable $r$ used above and $z$ used in eq.(\ref{metrcond}) we see that 
\be
\label{relrzxx}
\lambda_2' F_{ry}'=-{1\over a^2} \lambda_2'  F'_{zy}= -{1\over a^2} [{j_y'\over 4} -  \lambda_{10}' E_{y}']
\ee
where on the rhs we have also used eq.(\ref{bjy}). 
 
To complete the calculation we  need to express $T_{tx}$ in terms of boundary  theory coordinates. This requires 
us to multiply the rhs of eq.(\ref{expt}) by a factor of $a$. After doing this we get in the boundary  theory
\be
\label{btttx}
T_{tx}=({1\over i \omega})[-4 (Q_e'-\lambda_{10}'Q_m') E_x'-j_y'Q_m'+4 \lambda_{10}'Q_m' E_y']
\ee
 
\noindent Finally using the relation 
\be
\label{rethermo}
Q_x=T^{tx}-\mu J^x=-T_{tx}-\mu J_x=  T \alpha_{xx} E_x +  T \alpha_{yx} E_y
\ee
gives
\be
\label{alphaxx}
\alpha'_{xx}={(n'-4 \lambda_{10}' Q_m') \over i \omega T} +{Q_m'\over i \omega T} \sigma'_{yx} - {\mu' \over T}  \sigma'_{xx}
\ee
\be
\label{alphaxy}
\alpha'_{xy}={1\over i \omega T}[\sigma'_{yy} Q_m'-4 \lambda_{10}' Q_m']- {\mu' \over T}  \sigma'_{xy} 
\ee
where we have used the relation $n'=4 Q_e'$. 
By symmetries $\alpha'_{yy}=\alpha'_{xx}, \alpha'_{yx}=-\alpha'_{xy}$.

We have considered the action eq.(\ref{acsl}) in the analysis above, but it is easy to see that the relations eq.(\ref{alphaxx}), eq.(\ref{alphaxy})
stay the same for the  more general case 
\be
\label{genactcond}
S=\int d^4x\sqrt{-g}[R-2\Lambda - 2 (\partial \phi)^2 -h(\phi) (\partial \lambda_1)^2 - \lambda_2 F^2 - \lambda_1 F \tilde{F}], 
\ee
with  $h(\phi)$ and $\lambda_2$  being   general functions of $\phi$. 

The results above are quite analogous with those in \cite{HH}, which studied transport properties in the AdS Reissner-Nordstrom case.
It is instructive to compare the  cases with and without a  dilaton-axion.
Consider first the  purely electric case.   
We have seen earlier that  the thermodynamics in the extremal limit for the cases with and without a dilaton are  quite different, since the 
entropy vanishes in the presence of a dilaton.  
Despite this difference, we have also seen that 
the electric conductivity at both small and large frequency and small and large temperature qualitatively agree. 
In this subsection, we find  that the relation between the thermoelectric and electric conductivities is  essentially the
 same in the two cases.
Thus, the thermoelectric conductivity also agrees qualitatively in the two cases.  
Once a magnetic field is turned on, in the presence of an axion the thermodynamics of the 
 extremal situation continues to behave differently from the 
extremal Reissner-Nordstrom case, with vanishing entropy, while we saw in the previous subsection that  the electrical conductivity   is  still quite similar. 
Here we see that  the thermoelectric
conductivity gets additional contributions due to the presence of the axion, but these only affect the imaginary part and not the dissipative
real part at non-zero frequency. Thus, the thermoelectric conductivity continues to be quite similar.

\subsubsection{Thermal conductivity}
Next we turn to the thermal conductivity. It is easy to see using a Kubo formula that the thermal conductivity $\kappa_{ij}$ is related to the 
retarded two-point function of the heat current $Q_i$ \cite{HH}, 
\be
\label{deftcond}
\kappa_{ij}=-{\langle Q_i,  Q_i \rangle \over i \omega T}.
\ee
Using the definition of $Q_i$ eq.(\ref{dehc}) one then gets
\be
\label{mantp}
\langle Q_i, Q_j \rangle  =   \langle (T^{t}_i-\mu J_i),  - \mu J_j) \rangle + \langle T^t_i , T^t_j \rangle -\mu \langle J_i,  (T^t_j-\mu J_j ) \rangle -\mu^2  \langle J_i,  J_j \rangle \,.
\ee
Now it is easy to see from the rules of AdS/CFT that 
$$\langle (T^t_i-\mu J_i),  J_j \rangle =  \langle J_j,   (T^t_i-\mu J_i) \rangle$$
 so that the first and third terms on the rhs can
be related to each other. Further using the definition of thermoelectric and electric conductivity,
\be
\label{relconkubo}
\langle T^t_i-\mu J_i),  J_j \rangle =(-i \omega T) \alpha_{ij}, \ \  \langle J_i,  J_j \rangle =(-i \omega ) \sigma_{ij}
\ee 
then gives
\be
\label{finalkubkt}
\langle Q_i,  Q_j \rangle =i\omega \mu T (\alpha_{ij}+\alpha_{ji}) + i \omega \mu^2 \sigma_{ij} +  \langle T^t_i,  T^t_j \rangle.
\ee
As we will see in Appendix B
\be
\label{tttwo}
\langle T^t_i,  T^t_j \rangle = {\rho \over 2} \delta_{ij}
\ee
where $\rho$ is the energy density. 
Substituting the last few equations in eq.(\ref{deftcond}) then  finally gives the relation 
\be
\label{kappaij}
\kappa_{ij}=-\mu(\alpha_{ij}+\alpha_{ji})-{\mu^2\over T} \sigma_{ij}+{i \over 2 \omega T} \rho \delta_{ij}.
\ee

We note that this relation follows essentially  from the Kubo formula and is valid in general. 
For the case where there is no magnetic field we get from eq.(\ref{alphaxx}) and eq.(\ref{kappaij})
\be
\label{wf}
Re(\kappa_{xx})=({\mu^2\over T}) Re(\sigma_{xx}).
\ee
This is a Weidemann-Franz like relation, and is analogous to those obtained in the non-dilatonic case studied in  \cite{HKMS,HH}.
At low temperature and frequency,
 we have seen in \S3 that  ${\rm Re}(\sigma)_{xx} \sim {T^2\over \mu^2}$, leading to a linear behaviour of thermal conductivity 
\be
\label{wftwo}
Re(\kappa_{xx})\sim T.
\ee

The derivation of eq.(\ref{tttwo}) is discussed in Appendix B. We note that the result in eq.(\ref{tttwo}) is independent of 
momentum, and is therefore a contact term. Often in AdS/CFT calculations such contact terms are simply discarded. We  do not delve into this 
 issue here any further   except to note that    \cite{HH}  discusses it  and  does subtract this term from the final answer.

\subsection{Disorder and power-law temperature dependence of resistivity}
So far we have neglected the effects of disorder.  In this subsection we 
 attempt to include some of these effects  and discuss  the resulting consequences. 
Disorder can  be incorporated in a phenomenological way 
by adding a small imaginary part to the frequency, following \cite{HKMS},
$\omega\rightarrow \omega + i/\tau$.
We  focus on the resulting effects on   electric conductivity in the discussion below.

To begin,  consider the purely electric case.
The conductivity, at small frequency,  is given by eq.(\ref{locond})
\begin{equation}
\label{lca}
\sigma_{xx}={C'T^2\over \mu^2}+i C''{\mu\over (\omega+i/\tau)},
\end{equation} 
with $\sigma_{xy}=0$. 
For very small frequencies, $\omega \ll 1/\tau$ the disorder will dominate the imaginary part of $\sigma_{xx}$
and we get,
\be
\label{sxlc}
\sigma_{xx}\simeq C''\mu \tau + {C'T^2\over \mu^2}.
\ee
The first term on the rhs is a Drude-like contribution to the conductivity which is  proportional 
to the relaxation time $\tau$. For small disorder,  $\mu \tau \gg 1$ and
 we see that first term on the rhs of eq.(\ref{sxlc}) is large \footnote{
$C''$ which is dimensionless is $O(1)$.}. In the theory without disorder $Im(\sigma_{xx})$ has a pole and 
 $Re(\sigma_{xx})$ has a corresponding delta function at $\omega=0$. We see  from eq.(\ref{sxlc}) that after 
adding disorder, the pole and the delta function 
have both disappeared as expected, leaving a  large, but finite,  Drude-like contribution in $Re(\sigma_{xx})$.

Now consider the purely magnetic case obtained by carrying  out an $S$ transformation,
eq.(\ref{tsdef}) on the purely electric case.
Since $\tilde{a}=d=0$ we see from eq.(\ref{sigmayx})
 that $\sigma'_{yx}=0$ and since $c=1$ from  eq.(\ref{sigmaxx}) that  the resistivity,
\be
\label{lcb}
\rho_{xx}'={1 \over \sigma'_{xx}}={\sigma_{xx} \over 16}.
\ee
Thus the large Drude-like contribution in $\sigma_{xx}$ discussed above turns into a large resistivity in the magnetic
case,  scaling with 
the relaxation time $\tau$. In addition we see that the resistivity  now grows as $T^2$ with increasing temperature. 

The $S$ duality transformation is also  a symmetry of the purely dilaton theory, which does not have an axion, 
for all values of the coupling $\alpha$ defined in  eq.(\ref{ouraction}).
 Thus our results apply to these cases as well. 
More generally, see e.g.  \cite{Kiritsis},  once an additional potential is added for the dilaton-axion, one expects
that  the conductivity in the purely electric case can vary with temperature in  ways different from the $T^2$ dependence
we have found. This will then result in a different dependence for the resistivity in the purely magnetic case. 
In particular, we expect that one can obtain a linear dependence $\rho_{xx}\sim T$ reminiscent of 
strange metal behaviour in this manner. 



\subsection{$SL(2,R)$ and $SL(2,Z)$ in the boundary theory}

It is natural to ask about how the $SL(2,R)$ symmetry is implemented in the boundary theory. 
The gauge symmetry in the bulk corresponds to a global symmetry in the boundary. To implement the $SL(2,R)$ in the boundary one needs 
to gauge this global symmetry \cite{Witten2}. This is because, starting with a state which carries only electric charge in the bulk, one 
gets after a general $SL(2,R)$ transformation a system 
with both electric charge and a  magnetic field. Now, the magnetic  field corresponds to a non-normalisable deformation and therefore requires a
change in the boundary Lagrangian. Once the global symmetry is gauged in the boundary theory, there is a boundary gauge field $a^{\mu}$,
 and the required change in the boundary Lagrangian  can be identified as  turning on a background magnetic field. 

\subsubsection{$T_b$}

The $SL(2,R)$ symmetry is generated by the two elements $T_b$ and $S$ discussed in (\ref{tbdef}) and (\ref{tsdef}).
Under $T_b$ the axion shifts, $\lambda_1 \rightarrow \lambda_1 + b$. It is natural to identify this with a change in the coefficient 
of the Chern-Simons term for the gauge field in the boundary theory \cite{Kraus}.  In fact, this cannot be the whole story.
The reason is that, even for abelian gauge fields, the Chern-Simons term must appear with a quantised coefficient
\cite{Witten2}.  In defining the Chern-Simons term on a three-manifold $\Sigma_3$, one chooses an extension of the gauge
field to a four-manifold $\Sigma_4$ with $\partial \Sigma_4 = \Sigma_3$, and writes
\begin{equation}
\label{csdef}
\int_{\Sigma_3} A \wedge dA = \int_{\Sigma_4} F \wedge F~.
\end{equation}
Of course, to avoid arbitrariness in the definition, (\ref{csdef}) must yield an answer which is independent of the choice of 
$\Sigma_4$ and the extension of the gauge field -- or more precisely, the action $S(A)$ should depend on this choice only up to shifts by
integer multiples of $2\pi$, so that $e^{iS}$ is invariant.  This condition leads to a precise quantisation of the coefficient of the
Chern-Simons term.

Now, this poses a mystery, because in our system the Hall conductance takes arbitrary rational values (once we relax the full $SL(2,R)$ symmetry to
the more realistic $SL(2,Z)$).     However, this does not require violation of the quantisation condition.  Rather, consider
a (toy, boundary) Lagrangian of the form
\begin{equation}
S =  {1\over 4\pi} \int \left( k  ~\epsilon_{\mu\nu\rho} A_\mu \partial_{\nu} A_\rho ~-~  {1\over 2\pi} a^{\mu} \epsilon_{\mu\nu\rho} \partial_{\nu} A_{\rho}\right)~.
\end{equation}
This is the sort of Lagrangian that one finds in effective field theory descriptions of the quantum Hall effect; $A^\mu$ is to be identified with the
``emergent" gauge field (so the electromagnetic current is $J_{\mu} = {1\over 2\pi} \epsilon_{\mu\nu\rho} \partial^{\nu}A^{\rho}$) and $a^\mu$ is the external
electromagnetic field.  Integrating out $A^\mu$, one finds an effective Lagrangian for $a^\mu$ which gives fractional Hall conductance, and is roughly
a Chern-Simons theory at level $1/k$ \cite{Zee}.  Identifying $J^\mu$ with the global current in our boundary theory, and $a_{\mu}$ with the boundary
gauge field, we see how ``effective" fractional Hall conductances can arise in a theory with well-quantised Chern-Simons terms.  The generalisation
to describe arbitrary fractional quantum Hall states is discussed in, for instance, \cite{Zee}.

\subsubsection{$S$}

The $S$ transformation is more complicated. It action in the boundary theory has been discussed in \cite{Witten2}. 
In $2+1$ dimensions (at least  in the absence of charged matter) the  gauge field  $a^{\mu}$ is dual to a scalar $\phi$. 
The dual scalar theory has  a global symmetry, $\phi \rightarrow \phi+c$. 
The S transformation requires gauging this global symmetry 
and turning on a magnetic field for the resulting dual gauge field. 
This prescription for implementing $S$ also 
 roughly agrees with the discussion in \cite{HKMS} in which the $S$ duality acts by turning electrically charged 
particles into vortices. Electrically charged particles of the gauge field $a^{\mu}$
are vortices under the global symmetry  for $\phi$. Gauging the global symmetry corresponds to turning on a gauge field which couples (via local couplings) to these
vortices. 

In  the bulk, $SL(2,R)$ invariance means that the  theory comes back to itself with a 
different electric and magnetic field and altered dilaton-axion. This means in the boundary  too,  starting with the gauge theory containing the gauge field $a^{\mu}$ and carrying out the $SL(2,R)$ transformation should give  back the same gauge theory with the new magnetic field and  couplings 
corresponding to the new dilaton-axion and in a state with the new charge.

\subsubsection{$SL(2,R)$ vs $SL(2,Z)$}

In string theory, one does not expect that the $SL(2,R)$ symmetry is exact.
Instead it will 
be broken to an $SL(2,Z)$ subgroup 
generated by the elements $T_{b=1}, S.$ It is this $SL(2,Z)$ subgroup which should be an  symmetry (in the sense described above)
 of the boundary theory as well. 
The breaking of $SL(2,R)$ to $SL(2,Z)$  occurs due to stringy or quantum corrections in the supergravity action; it can also be understood as
being related to charge quantisation. In any case, at the level of bulk solutions, if the supergravity approximation we  are working with  here
is good, at large values of the charges the supergravity will have an approximate $SL(2,R)$ symmetry and the approximation we make discussing the full
$SL(2,R)$ is a good one. This means  our conductivity and thermodynamic calculations
using the $SL(2,R)$ to relate situations with different electric and magnetic charges should be accurate, and the 
$SL(2,R)$ transformations in the boundary theory should be approximately valid.    One can always restrict consideration to $SL(2,Z)$ transformations
acting on the electrically charged brane with minimal charge, to get a more accurate picture.

\FIGURE[ht!]{\includegraphics[scale=0.8]{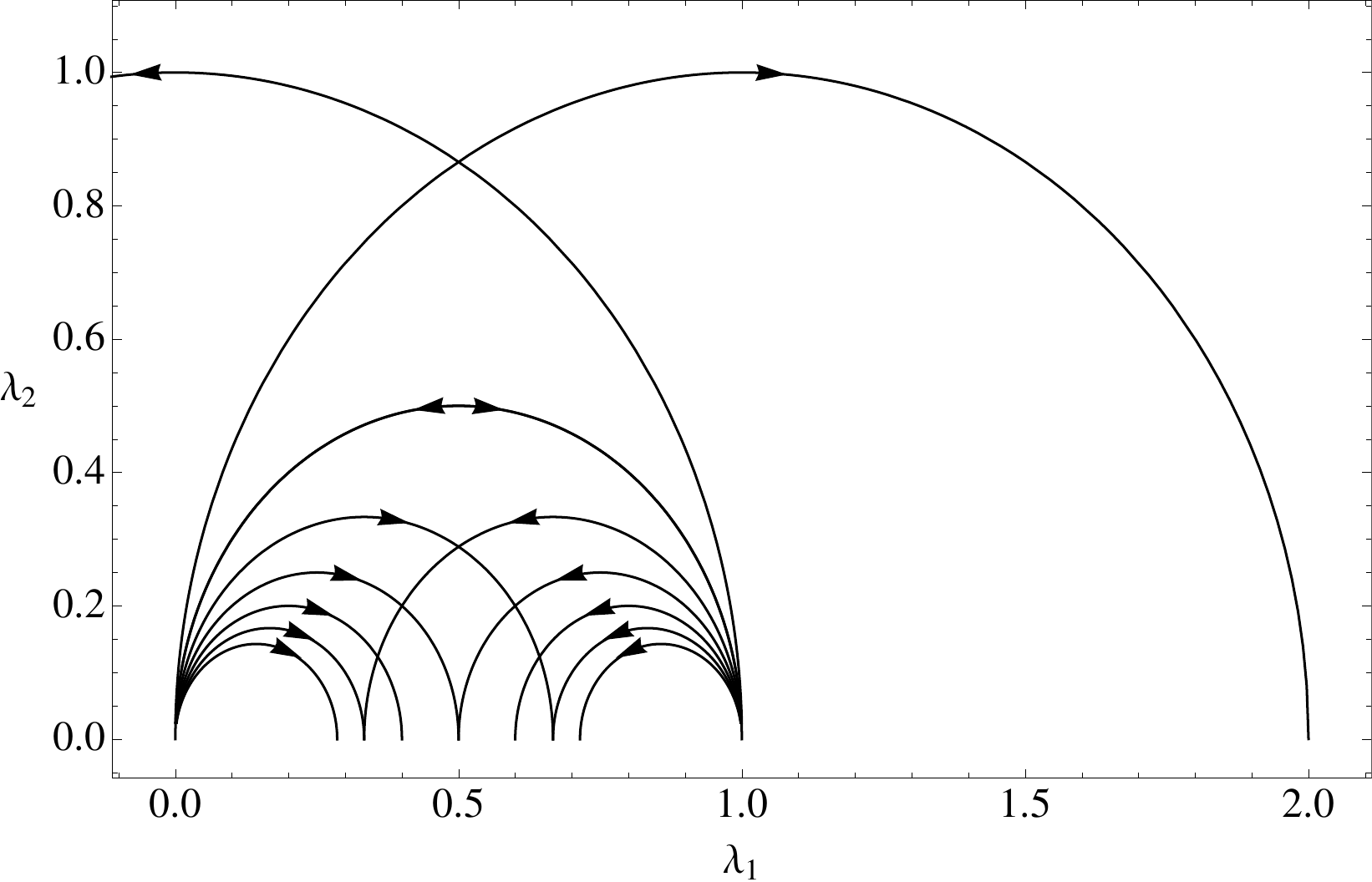}
\caption{Attractor flows for various ${\cal O}(1)$ values of $Q_e/Q_m$, in the case without a bare scalar potential.  The trajectories run from $r=100$ to $r=10^{-5}$, along the direction
indicated by the arrows.  The various semi-circles arise for values of $Q_e/Q_m$ varying from 2 (for the largest one) to 2/7 (for the smallest one); the incomplete semi-circle would hit the real axis at $\lambda_1=-1$ if we extended the figure.  The initial
value of the axion is very close to 0, but flows which would go to the left of the vertical axis have been shifted (via the $T_{b=1}$ transformation $\lambda_1 \to \lambda_1+1$) to appear in the figure.  The initial dilaton value is fixed by
an appropriate $SL(2,R)$ mapping of the value of $\lambda_2\vert_{r=100}$ in the electric solution.}
\label{fig.1}
}

\section{Attractor behaviour in systems with $SL(2,Z)$ symmetry}

 In this section, we discuss the structure of attractor flows in the dilaton-axion plane.  We begin by discussing the flows governed by the action
 $(5.1)$, and then consider a more general action which includes an $SL(2,Z)$ invariant potential for the dilaton-axion.   The main feature of interest
 here is that the $SL(2,Z)$ symmetry acts to relate different attractor flows to one another; in the field theory, this would mean that different RG 
 trajectories are related by the modular group.  In the system without a potential, the endpoints of the flows have rational $\sigma_{xy}$ and vanishing
 longitudinal conductivity, while in the system including a potential, we find (at fixed charges) different basins of attraction for distinct attractors:
 some at rational values of $\lambda_1$ and strong coupling, and others at $\lambda \simeq i$.
 
 In addition to the intrinsic interest of the subject, we are motivated to study the action of $SL(2,Z)$ on these flows because $SL(2,Z)$ (or more properly,
 its subgroup $\Gamma_{0}(2)$) has been
 argued to organise the phase diagrams of real systems of charged particles in background magnetic fields.  Discussions in the context of the
 fractional quantum Hall system can be found in \cite{Wilczek, Lutken, Zhang, Burgess}, and a nice review appears in
 \cite{Dolan}.  Needless to say, it would be very interesting to modify our system to give incompressible phases and analogues of Hall plateaux, 
 but we do not pursue this here.  Discussions of holography and the quantum Hall system can be found in \cite{Kraus,Kraus2,Fujita, Kraus3,Bak, Bergman}.
  
Before proceeding, we should emphasise that there is an obvious difficulty with controlling the RG flows of greatest interest in our system.
With a magnetic field turned on, 
the IR-attractor lies along the real axis in $\lambda$,
at strong coupling.  To the extent that one can trust the analysis it is attractive for both the dilaton and axion directions.
More correctly, close enough to the fixed point, supergravity breaks down and corrections  would have to be included to study the nature of the RG flow in more
detail.  In this section, we will simply take the attractor flows at face value.

 \subsection{Attractor flows in the $SL(2,R)$ invariant theory} 
 
 One wide class of attractor flows in the $SL(2,R)$ invariant case are easily determined, as follows.  The flows in the original electric solutions of \cite{GKPT} 
 are extremely simple, involving logarithmic variation of the dilaton (running to weak coupling at the horizon).   Using the $SL(2,R)$ transformation properties
 of the dilaton-axion $(5.5)$, one can translate these dilaton trajectories into more non-trivial dilaton-axion trajectories, governing the flow to dyonic black holes in
 the extremal limit.   A plot displaying these trajectories for various ${\cal O}(1)$ ratios of $Q_e/Q_m$ appears in Figure~\ref{fig.1}.   It is clear from the nature of the $SL(2,R)$
 duality, which relates the axion to $e^{-2\phi}$, that the axion is attracted to its fixed-point value in a power-law manner.
 
 All of the fixed points in this case lie on the real $\lambda$ axis, with rational values of $\lambda_1$ (and hence $\sigma_{xy}$) and vanishing $\sigma_{xx}$.  
 Because of the extreme value of the dilaton at infinity, these states are also incompressible.
 This is
 happily rather similar to the flows in the quantum Hall system, but the underlying physics of our charged fluid is quite different.
  


\subsection{Attractor flows in the presence of a potential which breaks $SL(2,R)$ to $SL(2,Z)$}
 
 In the flows of interest in more realistic systems, there are also RG fixed points at other fixed points of subgroups of $SL(2,Z)$ in the upper half plane.   To find additional
 fixed points in our case, we must add a bare scalar potential.  This modifies the effective potential governing the attractor flow as in e.g. (2.37)-(2.39) of \cite{GKPT}.  
 Here, we discuss the most natural class of $SL(2,Z)$-invariant potentials (which do, however, break $SL(2,R)$ to $SL(2,Z)$).
 
 Our dilaton-axion kinetic terms can be derived from the K\"ahler potential
 \begin{equation}
 K = - {\rm log}\left(-i (\lambda - \overline{\lambda})\right)~.
 \end{equation}
 It is then natural to try and derive an $SL(2,Z)$-invariant potential by choosing an appropriate superpotential $W$ and using the supergravity formula
 \begin{equation}
 V = e^{K} \left( G^{\lambda\overline{\lambda}} D_{\lambda}W \overline{D_\lambda W} - 3|W|^2\right)~.
 \end{equation}
 Here, $G$ is the K\"ahler metric derived from $K$, and $W$ must transform as a section of a particular line bundle over the dilaton-axion moduli space,
whose first Chern class is determined by $K$ \cite{Bagger}.

In fact, the superpotentials which are allowed by this requirement, and satisfy in addition the physically sensible condition of vanishing at weak coupling (where we
know that potentials slope to zero in realistic string compactifications), have been discussed extensively in earlier literature \cite{Moore}.   
Defining
\begin{equation}
q = e^{2\pi i \lambda}
\end{equation}
they take the rough form
\begin{equation}
W = {1\over \eta(q)^2} ~f(j(q)),
\end{equation}
 where $j$ is the famous $j$-invariant modular function 
 \begin{equation}
 j(q) = {1\over q} + 744 + 196884 ~q + \cdots
 \end{equation}
 and 
 \begin{equation}
 \eta(q) = q^{1/24}~ \Pi_{i=1}^{\infty} \left(1-q^n\right).
 \end{equation}
 Because $j$ has a pole at weak coupling $j(q) \sim {1\over q}$, one should choose $f$ to die quickly enough as $j \to \infty$ near weak coupling to satisfy the
 requirement $V \to 0$ as $g_s \to 0$.
 
 One simple choice \cite{Moore}, which suffices to add an interesting new critical point near $\lambda \simeq i$ in our attractor flows, is the choice
 \begin{equation}
 W = {1\over \eta(q)^2} {1\over j(q)^{1/3}}~.
 \end{equation}
 With this choice, the scalar potential takes the form
 \begin{equation}
 \label{hugemess}
 V(\lambda) = e^{-\pi \lambda_2} {\vert \tilde \eta(q) \vert^{12} \over {\lambda_2 \vert E_4(q)\vert^4}} \left(
 \vert E_4(q) + {\pi \over 3} \lambda_2 (3 E_2(q) E_4(q) - 4 \tilde E_4(q)) \vert^2 - 3 \vert E_4(q)\vert^2 \right)~.
 \end{equation}
 Here, $E_2$ and $E_4$ are the Eisenstein functions
 \begin{equation}
 E_2(q) = 1 - 24 \sum_{n=1}^{\infty} {nq^n \over {1-q^n}},~~E_4(q) = 1 + 240 \sum_{n=1}^{\infty} {n^3 q^n \over {1-q^n}}
 \end{equation}
 and
 \begin{equation}
 \tilde E_4(q) = {3\over 2\pi i} \partial_{\tau} E_4(q)~,~~\tilde \eta(q) = q^{-1/24} \eta(q).
 \end{equation}

 \TABLE{
\centering
$
\begin{array}{|c|c|c|c|}
\hline & & & \\[-2.2ex]
 \frac{Q_e}{Q_m} & \lambda _{1_*} & \lambda _{2_*} & b_h \\[0.7ex] \hline & & & \\[-2.ex]
 2 & 6.2\times 10^{\text{-3}} & 1.004 & 0.76 \\[1.5ex]
 1 & 7.7\times 10^{\text{-3}} & 1.000 & 0.86 \\[1.5ex]
 \frac{2}{3} & 7.1\times 10^{\text{-3}} & 0.997 & 0.97 \\[1.5ex]
 \frac{1}{2} & 6.2\times 10^{\text{-3}} & 0.996 & 1.08 \\[1.5ex]
 \frac{2}{5} & 5.3\times 10^{\text{-3}} & 0.995 & 1.19 \\[1.5ex]
 \frac{1}{3} & 4.6\times 10^{\text{-3}} & 0.994 & 1.29 \\[1.5ex]
 \frac{2}{7} & 4.1\times 10^{\text{-3}} & 0.994 & 1.38
\\[1ex] \hline
\end{array}
$
\caption{Precise locations of the new attractor points near $\lambda \simeq i$, and values of $b_h$ (the value of $b(r)$ at the horizon), for various values of $Q_e/Q_m$.}\label{tab.1}
}

 The main feature of interest in the potential (\ref{hugemess}) for us, is that it has a minimum at $\lambda = i$ (in addition to runaway minima close to $i\infty$, and
 a singularity at $e^{2\pi i \over 3}$).  It will turn out that at least in some cases, this translates into a critical point in the full effective attractor potential for dyonic
 dilaton black holes, when the bare potential (\ref{hugemess}) is added to the action $S$ in $(5.1)$.   
 
 To see this, let us remind ourselves of the class of attractors described in \cite{GKPT}, eqns. (2.37)-(2.39).  In the presence of a bare potential $V_1$, in addition
 to the effective potential $V_{eff}$ arising from the gauge kinetic function in a charged black hole background, one can find $AdS_2 \times R^2$ attractor points
 if there exist a $b_h$ and $\lambda^*$ such that:
 \begin{equation}
\partial_{\lambda} V_{eff}(\lambda^*) + b_h^4 ~\partial_{\lambda} V_1(\lambda^*) = 0~,
\end{equation}
\begin{equation}
\left( {3\over L^2} - V_1(\lambda^*)\right) ~b_h^4 = V_{eff}(\lambda^*)~.
\end{equation}

Consider now a family of actions:
\begin{equation}
S_{\epsilon} = S - \epsilon \int d^4x  ~\sqrt{-g} V(\lambda)~
\end{equation}
with $V$ the modular-invariant potential discussed above.
That is, we modify the action in $(5.1)$ by adding the potential (\ref{hugemess}), with strength $\epsilon$ (or, in terms of the discussion above, $V_1 = \epsilon V(\lambda)$).

Suppose (as is actually the case) that the potential $V(\lambda)$ has an AdS minimum at some point in field space.
It is clear that at sufficiently large values of $\epsilon$, at any finite $b_h$, the potential term will dominate over the gauge kinetic function in determining
the properties of an attractor geometry.  Then, the first equation above just becomes the condition for a critical point of $V(\lambda)$, while the second equation
can be solved for $b_h$.  Of course at any finite $\epsilon$, the attractor will be slightly shifted from the minimum of $V(\lambda)$ by the effects of the gauge
kinetic function.\footnote{It is important that the minimum of $V$ under discussion be an AdS minimum; if it is a dS minimum, then for sufficiently large
$\epsilon$ the overall effective cosmological term changes sign in the vicinity of the minimum, and our discussion would be radically modified.}

On the other hand, it is also clear that for small $b(r)$, the gauge kinetic contributions to the attractor potential dominate over the contributions of the bare
potential.  This is because one of them scales like $1/b^2$, while the other scales like $b^2$.  Therefore, one can expect to find our Lifshitz-like attractors with
$b \to 0$ at the horizon, even in the presence of the bare potential $V$.\footnote{This would not necessarily be true if $V$ diverged sufficiently strongly at
the attractor point, but in fact $V$ vanishes along the real $\lambda$ axis -- this was actually one of our conditions for reasonableness of the potential, since the real axis is $SL(2,Z)$ dual
to weak coupling.}

In fact, we find that at moderate values of $\epsilon$, the resulting system exhibits multiple attractor points at fixed charges.  That is, for reasonable choices
of $\epsilon$, and a fixed $Q_e, Q_m$, one finds both the attractor point at $\lambda = Q_e/Q_m$, and an attractor very close to $\lambda = i$.
We give some representative values of the moduli at the new attractor point near $\lambda_* \simeq i$ for various ${\cal O}(1)$ values of $Q_e/Q_m$ and
for $\epsilon = 144$ in Table~\ref{tab.1}.\footnote{ $\epsilon = 144$ is a convenient choice, because Mathematica naturally defines the $j$-function in a way that differs
from the standard definition by a factor of 1728.}  The standard attractors at $\lambda = Q_e/Q_m$ also exist in all of these cases, and the $\lambda$-plane is
divided into different domains which flow to one attractor or the other.

\section{Attractor behaviour in more general system without $SL(2,R)$ symmetry}

In this section, we study a  more general theory which does not have $SL(2,R)$ symmetry. 
The action we study has one parameter $\alpha \neq -1$, 
\bea
\label{generalaction}
S = \int d^4x \sqrt{g} \left( R -2 \Lambda - 2 \left( \partial \phi \right)^2 - \frac{1}{2} e^{4 \phi} 
\left( \partial  \lambda_1\right)^2 
- e^{2 \alpha \phi} F^2 - \lambda_1 F \tilde F \right)
\eea
We will analyse the attractor mechanism for  dyonic black branes in this theory.\footnote{Of course other parameters in the 
action eq.(\ref{generalaction}) could have also been varied from  their values in the $SL(2,R)$-invariant case. We do not carry out a full analysis
of the resulting set of theories here, but the limited class we do study already exhibit rather interesting phenomena.}

The effective potential is now given by
\bea
\label{effectivepot}
V_{eff}(\phi, \lambda_1) = e^{- 2 \alpha \phi} \left(Q_e - \lambda_1 Q_m \right)^2 + e^{2 \alpha \phi} Q_m^2,
\eea
where $Q_e,Q_m$ are the electric and magnetic charges. 
The extremum of the potential arises at 
\be
\label{attgen}
\lambda_1=\lambda_{1*}={Q_e\over Q_m}, \ \ e^{2\alpha \phi}\rightarrow -\infty.
\ee
We work in the  coordinate system eq.(\ref{ansatz}) below. 
If the axion takes its attractor value $\lambda_{1*}$ at $r\rightarrow \infty$,  it is constant everywhere and the resulting
 solution  is that of a purely  magnetically charged dilatonic brane. This has a 
 near horizon metric given in eq.(\ref{nhmetricl})
and the near-horizon dilaton
\be
\label{dilpmag}
\phi=K\log(r), 
\ee
with the constants  $C_2,\beta, K$ taking values given in eq.(\ref{constants}). 

To investigate if this magnetic solution is an attractor, we  take the asymptotic value of the axion at infinity to 
be slightly different 
from its attractor value and study the resulting solution. As we will see below,  
in the ranges $\alpha>0$ and $\alpha \le -1$ we find attractor behaviour, with the axion settling down to its attractor 
value  exponentially rapidly in $r$ (except for the special case $\alpha = -1$ discussed in \S6, \S7, where the attractor is power-law in nature).  
In the range $-1<\alpha<0$ we find  that there is no attractor behaviour.
Instead, starting with a value for the axion at infinity which is slightly different from its attractor value,
one finds that the 
 solution increasingly deviates from the 
purely magnetic case for small enough $r$. We have not been able to find the end point of the attractor flow in this case.


\subsection{Attractor behaviour for $\alpha>0$, $\alpha<-1$}
The axion equation of motion is 
\be
\label{axeom}
\partial_r(e^{4\phi} a^2b^2\partial_r\lambda_1)={4e^{-2\alpha\phi}Q_m^2  \over b^2} (\lambda_1 -\lambda_{1*})
\ee
Putting in the solution for $\phi, a^2,b^2$ in the near horizon region of the purely magnetic case  gives
\be
\label{nhformge}
\partial_r(r^{4K+2\beta+2}\partial_r \lambda_1)={D  \over r^{2\beta+2\alpha K}}(\lambda_1 - \lambda_{1*})
\ee
where $D>0$ is a constant. 

Define the variable $x$ as
\be
\label{defax}
x={1\over |4K+1+2\beta|} {1\over r^{4K+1+2\beta}}
\ee
In terms of $x$ eq.(\ref{nhformge}) becomes a Schr\"odinger-type equation,
\be
\label{stypee}
\partial_x^2 \lambda_1= \tilde{D} x^{-P} (\lambda_1-\lambda_{1*})
\ee
where $\tilde{D}>0$ is a  constant and 
\be
\label{valP}
P={4K-2\alpha K + 2\over 4K+1+2\beta}.
\ee

By rescaling $x$ the constant  $\tilde{D}$ can be set to unity \footnote{This does not work for 
 the case $P=2$ which  arises when $\alpha=0, -1$. The $\alpha=-1$ case has $SL(2,R)$ invariance and has been extensively 
discussed above.
The $\alpha=0$ case needs to be dealt with separately because here the dilaton does not enter in the gauge kinetic 
energy or the  effective potential. }.   
To avoid notational clutter, we continue to refer to this rescaled variable as $x$ below. Also, to simplify things, 
we henceforth take $(\lambda_1-\lambda_{1*})\rightarrow \lambda_1$, i.e., from now on we  
  use $\lambda_1$ to denote the deviation of the axion from its attractor value. This gives
\be
\label{sttwo}
\partial_x^2\lambda_1=x^{-P}\lambda_1
\ee

There are two separate  cases of interest. 

\subsubsection{Case A}
The first case arises when  
\be
\label{caseaconda}
4K+1+2\beta>0
\ee
Here we see from eq.(\ref{defax}) that 
$x\rightarrow \infty$ as $r\rightarrow 0$. 
When 
\be
\label{caseacondb}
P<2
\ee
a solution to eq.(\ref{sttwo}) can be found in the WKB approximation.
It is of the form
\be
\label{wkba}
\lambda_1\sim e^{-S},
\ee
with
\be
\label{valssec7}
 S={x^{1-P/2}\over 1-P/2}. 
\ee
We see that as $x\rightarrow \infty,   S\rightarrow \infty$ and  $\lambda_1\rightarrow 0$, so the   axion goes 
to its attractor value in the near horizon region exponentially rapidly. 
In finding the solution we have neglected the backreaction of the axion on the other fields; this is now seen to
 be a self-consistent approximation. Since the other fields vary in a power law fashion with 
$r$, the backreaction of the axion on them is small.

Substituting for the constants from eq.(\ref{constants}) in the conditions eq.(\ref{caseaconda}) eq.(\ref{caseacondb}), we find that the solution eq.(\ref{wkba}) is valid in the range
\be
\label{rangeaa}
\alpha>0, \ \ {\rm or } \ \ \alpha<-2.
\ee
 
\subsubsection{Case B}
The second case arises when  
\be
\label{casebconda}
4K+2+2\beta<0
\ee
Now the variable $x \rightarrow 0 $ as $r\rightarrow 0$. 

A solution to eq.(\ref{sttwo}) can be found in the WKB approximation  when 
\be
\label{casebcondb}
P>2.
\ee 
It is again of the form given in  eq.(\ref{wkba}), with $S$  being
\be
\label{valscaseb}
S={x^{1-P/2}\over P/2-1}.
\ee
The conditions, eq.(\ref{casebconda}), eq.(\ref{casebcondb}) are valid when $\alpha$ lies in the range
\be
\label{rangecaseb}
-2<\alpha<-1.
\ee

\subsection{No attractor when $-1<\alpha<0$}
Our discussion above left out the region $-1<\alpha<0$.
In this region, we will see below that  there is no attractor behaviour. 

First, consider the case when $4K+1+2\beta>0$ and $P>2$, which corresponds to $-2/3<\alpha<0$. 
In this case, we see from eq.(\ref{defax}) that 
 $x\rightarrow \infty$ in the near horizon region where $r\rightarrow 0$. 
As discussed in appendix C there is only one  solution to the axion equation which does not blow up  
as $x\rightarrow \infty$. 
It takes the form 
\be
\label{natta}
\lambda_1=c_0+c_1 x^p, \\ p < 0 ~.
\ee
Here $c_0,c_1$ are two non-vanishing constants.  
Since $c_0$ does not vanish in this solution  $\lambda_1$ does not vanish as $x\rightarrow \infty$ and 
thus the axion does not reach its attractor value. 

Next consider the case when $4K+1+2\beta<0$ and $P<2$, which corresponds to $-1<\alpha<-2/3$. 
Here $x\rightarrow 0$, when $r\rightarrow 0$. In this case there is a solution  in which 
the axion attains its attractor value as $x\rightarrow 0$. As discussed in appendix C this takes the form, for small $x$,
\be
\label{smallxform}
a=c_1 x +c_2 x^p, \\ p > 1 ~.
\ee
 However, since  the approach to the attractor value is a power-law in $x$ and thus in $r$, one now finds 
that the resulting back-reaction of the axion in the equations of motion for the other fields 
cannot  be neglected, and in fact in some cases dominates over the other contributions. 
Thus, again, the resulting solution will deviate significantly from the purely 
magnetic case, leading to a loss of attractor behaviour.
 
In this last  case especially, one might hope to find a fully corrected solution which represents the end point of the attractor flow, in which all fields behave in a power-law fashion near the horizon, and in which the back-reaction of the axion is 
completely incorporated. A reasonably thorough analysis, 
which we have included in Appendix D, however failed to find any  purely power-law solution 
of this kind.

\subsection{Comments}
Let us conclude this section with some comments. 

In the cases where we did get attractor behaviour above, we saw that the axion approached its attractor value exponentially rapidly in the near-horizon region. This exponential behaviour  is intriguing from the point of view of a dual field theory. 
The radial direction $r$ is roughly the RG scale in the boundary theory and  a    power-law dependence on $r$ of a field in the bulk
is related  to the  anomalous dimension of the corresponding operator in the boundary. 
In contrast an  exponential dependence, of the kind  we find here,  leads to a beta function for the dual operator  
in the boundary in which  the RG scale appears explicitly. 

The exponentially rapid approach also means that in cases where we do get attractor behaviour, 
 the black brane in the near-horizon region  can   be taken to be the purely magnetic  dilatonic brane
 up to small corrections. 
 This means the behaviour of the dyonic black brane at small temperature and frequency in these cases
 is given by that of the  dyonic brane with the asymptotic axion set to its attractor value, 
 up to small corrections. For example, from eq.(\ref{valdb}) we see that when $\lambda_{1\infty}=\lambda_{1*}$ the $SL(2,R)$ matrix element $d$ vanishes. The conductivity can then 
be read off from eq.(\ref{condayx}), eq.(\ref{condxx}) keeping this in mind. 
Similarly the thermoelectric and thermal transport coefficients can also be found easily from 
eq.(\ref{alphaxx}), eq.(\ref{alphaxy}).

\section{Concluding comments}

We have analysed charged dilatonic branes in considerable detail in this paper, focusing on their thermodynamics and especially their 
transport properties. 
Our results show that many of the transport properties are quite similar to those of the Reissner-Nordstrom case.  This is true despite the fact 
that the Reissner-Nordstrom and dilaton cases differ significantly in their thermodynamics: while the Reissner-Nordstrom brane has a macroscopic
ground-state entropy,  the 
  dilatonic black brane
 has vanishing entropy at extremality.
 
More concretely,  in \cite{GKPT} it was already noted that the optical conductivity at zero temperature and small frequency
 has the  behaviour $Re(\sigma) \sim \omega^2$, and this behaviour 
is independent of the parameter $\alpha$  which governs the dilaton coupling, eq.(\ref{ouraction}). 
In particular, it is the same as in the Reissner-Nordstrom 
case which has $\alpha=0$ \cite{Leigh, Paulos}.  In this paper we find something analogous for the DC conductivity 
at  small temperature,
  which goes like \footnote{There is an additional delta function strictly at $\omega=0$.} $Re(\sigma)\sim T^2$, and is 
 independent of $\alpha$ again. 
In the presence of a magnetic field, the DC Hall conductivity is $\sigma_{yx}= {n\over B}$, where $n,B$ are the electric charge density and the magnetic field, while the DC longitudinal conductivity vanishes,  as required by Lorentz invariance. The DC Hall conductance is related to the 
attractor value of the axion.  
In more detail, the frequency dependence  
  fits the  form derived from  general considerations of relativistic magnetohydrodynamics in 
\cite{HKMS}. These features in the presence of  a magnetic field,  being general in their origin, also agree with the Reissner-Nordstrom case. We also found that the  thermoelectric and the thermal conductivities of the dyonic case satisfy Weidemann-Franz like relations which relate them to their electrical conductivity. In this respect too then the 
dyonic system behaves in a manner quite analogous to the Reissner-Nordstrom case.\footnote{ It is worth 
pointing out that, in contrast to these similarities,
the viscosity of a near-extremal dilaton-axion system is much smaller than in the Reissner-Nordstrom case. 
In both cases the famous relation ${\eta/ s}={1 / 4\pi}$ \cite{son2} is satisfied. However, the 
vanishing entropy of the extremal dilaton-axion system makes its viscosity much smaller.}

The overall picture,  then, is that  the charged dilatonic brane behaves  like  a charged plasma.  
The  electrical conductivity, which is suppressed at small temperature and grows like $T^2$, suggests that strong repulsion prevents the transmission of electric currents in this system. 
The spectrum is not gapped in the conventional sense above the ground state, since this would lead to a conductivity vanishing exponentially 
quickly at small temperature.  Rather, the system has a ``soft" gap, resulting in a power-law vanishing as $T \to 0$.\footnote{Strictly speaking, our calculations break down at extremality, so these comments apply for temperatures much smaller than the chemical potential, but not very close to zero. The precise condition 
can be obtained using reasoning analogous to eq.(\ref{condnhaa3}) in the magnetic case.}
It should be pointed out that the entropy density $s$  also scales in a power-law fashion as $s\sim T^{2\beta}$,  and since $\beta<1$, it 
decreases more slowly near extremality (as $T \to 0$) than the charge conductivity. This makes physical sense: only some fraction of all the degrees of freedom 
can carry  charge and contribute to electrical conductivity.

A case we investigated in considerable detail was the one  with  an $SL(2,R)$ symmetry. Here, the complex conductivities
$\sigma_{\pm}$ transform like the dilaton-axion under an $SL(2,R)$ transformation. Once quantum corrections to the bulk action are included (or charge quantisation is imposed), one expects this
symmetry to be broken to an $SL(2,Z)$ subgroup.   The transformation law for $\sigma_\pm$ is an 
elegant result, and one has the feeling that its full 
power has not been exploited in the discussion above.  Perhaps suitable modifications of the bulk theory, 
with an additional potential for the dilaton-axion  preserving the $SL(2,Z)$
symmetry and/or with disorder put in, might prove interesting in this respect. These modifications might lead to similarities  with systems 
 exhibiting the quantum Hall effect, and the 
transformation law of the  conductivity could then tie in with some of the  existing discussion  in this subject 
on RG flows between different fixed points 
characterised by the various  subgroups of $SL(2,Z)$ \cite{Wilczek, Lutken, Zhang, Burgess, Dolan}.  We briefly explored the addition of a
modular-invariant potential in \S7, but it seems likely that a deeper investigation of the case with potentials could be fruitful.  Such an
investigation, in the case of electrically charged dilatonic black branes, was recently undertaken in \cite{Cadoni, Kiritsis}.\footnote{We note that in theories where
the dilaton has an axion partner (e.g. supersymmetric theories), potentials which are exponentials in $e^{-2\phi}$ are more natural than simple exponentials in $\phi$, because of the axion
shift symmetry (which is typically valid to all orders in perturbation theory).  The potential
we investigated in \S7\ has this form near weak-coupling, but this has not been the case for most potentials investigated in the literature on this subject.}

We have not shown that the  dilaton-axion theories considered here can arise in string theory. However, the Lagrangians we consider are quite simple and generic, and as discussed above many of our  results  are  quite  robust. These facts suggest that an embedding in string theory 
  should be possible.   String embeddings of Lifshitz solutions have been described in \cite{Takayanagi, Joe, Koushik}, and simple generalisations of
  those ideas may well suffice to capture our geometries as well (since the near-horizon physics is governed by a Lifshitz-like metric).

Our main focus in this paper was on cases where there is no bare potential for the dilaton-axion (i.e., where it corresponds to an exactly marginal
operator in the dual field theory).    However, we did briefly discuss addition of a modular-invariant potential in \S7, and one would expect that
the models which arise in string theory would generically have some potential.  As we saw
in \S7, if this
 potential  has a sufficiently weak dependence on the moduli, our analysis will still go 
through with only small corrections (since the gauge kinetic function is favoured by extra powers of $1/b(r)$ as compared to the bare potential,
and $b$ becomes very small in the near-horizon region in our Lifshitz-like near-horizon geometries). 
Of course, in the landscape of vacua, one expects there to be many more theories  where the dilaton-axion dependence of the 
potential is not small. How different will these cases be? It is clear from the study of simple model cases, e.g. in \cite{Cadoni} and also in the
very thorough treatment of dilatonic branes in \cite{Kiritsis}, that
the exponents governing the power law dependence of the optical conductivity on frequency or the DC conductivity on temperature,
can be modified from the values we found by the presence of a potential.  Different power laws can also be found by considering $U(1)$ gauge fields
on probe branes in Lifshitz-like backgrounds \cite{Joe}.
So the precise exponents we have found are, unfortunately, not likely to be universal results for gravitational systems. However, the feature that these dependences are power-laws 
might itself be one of considerable generality. In gravitational systems, one expects that the far infrared of extremal branes is  
characterised by an attractor geometry, with an emergent scale invariance up to possible logarithmic corrections. As a result,
 the frequency and temperature dependences  should be governed by power-laws determined by the scaling dimensions of the operators of interest. 

We are not aware, at the moment, of condensed matter systems or model Hamiltonians which give rise to such a power-law behaviour 
in the conductivity.\footnote{The systems we are considering here do not have any disorder. In the presence of disorder such power laws are well known to arise 
\cite{Lee}. We thank P. Raichaudhuri and N. Trivedi for related discussions.}
It would be quite interesting to construct or find such examples, and attempt to relate their behaviour to the kinds of gravitational 
systems studied here.   

\bigskip
\centerline{\bf{Acknowledgements}}
\medskip

We thank A. Dabholkar, K. Damle, J. David, S. Hartnoll, P. Kraus, R. Loganayagam, M. Mahato, A. Maloney, S. Minwalla, M. Mulligan, R. Myers, P. Raichaudhuri,  S. Rey, S. Ross, S. Sachdev, A.  Sen,
E. Silverstein, A. Strominger, N. Trivedi and E. Witten for interesting discussions.
N.I. and S.P.T. acknowledges the warm hospitality of the LPTHE,  University of Paris, during a visit in which some of this work was done. N.I. also thanks RIKEN for their hospitality. S.P. thanks the Centre for High Energy Physics at the Indian Institute of Science, Bangalore for their kind hospitality during a visit where part of this work was done.
S.K. is indebted to the Kavli Institute for Theoretical Physics and the UCSB Physics Department for hospitality while the bulk of this research was carried out. 
He is also grateful to the theory groups at Caltech, CU Boulder, Harvard, IPMU, Johns Hopkins, Kentucky, McGill, the Mitchell Institute for Fundamental Physics, the Perimeter Institute, UBC 
and UCLA for hospitality and stimulating
discussions.
S.P. and S.P.T. acknowledge the support of DAE Government of India and most of all are grateful to the people of India 
for generously supporting research in string theory. 

\newpage

\appendix
\section{Appendix A}
In this appendix we carry out a more careful examination of the Schr\"odinger equation eq.(\ref{seb}) and show that  the  coefficient $a_1$ in 
eq.(\ref{soln}) is of 
order unity and not suppressed by a  power of $\omega$. 

The potential  $V(z)$ is given by  eq.(\ref{fformpot}). In the scaling region where $r\ll \mu$, after a 
suitable rescaling the metric and dilaton are 
given by eq.(\ref{nhmetric}), eq.(\ref{dilaton}),
with  coefficients given in eq.(\ref{constants}).  The constant $Q^2$  which appears in the potential  takes the value (eq.(2.12) of 
\cite{GKPT})
\be
\label{valq}
Q^2={6\over \alpha^2+2}.
\ee
We use the notation 
\be
\label{hatomega}
\hat{\omega}={\omega \over r_h}
\ee
below. 

At the horizon, where $a^2$ vanishes,  the potential has a first order zero and for 
\be
\label{regiona}
\hat{r}-1\ll 1
\ee
  it takes the form
\be
\label{nhpot}
V=A (\hat{r}-1), 
\ee
where  $A$ is a coefficient of order unity.
Also in this region the variable $\hat{z}$ eq.(\ref{hatz}) is given by 
\be
\label{varza}
\hat{z}=\int {d\hat{r}\over \hat{a}^2} \simeq {1\over B} \ln(\hat{r}-1)
\ee
where $B$ is again a coefficient of order unity.

We begin in the very near horizon region where 
\be
\label{rangea}
| \hat{r}-1| \ll {\hat{\omega}^2 \over A}.
\ee
In this region the potential is subdominant compared to the frequency in the Schr\"odinger equation and as a result, the   solution with the 
correct normalisation  to obtain the required flux  is
eq.(\ref{vchorizon})
\be
\label{solse}
\psi=e^{-i\hat{\omega}\hat{z}}
\ee
(there is an additional $e^{-iwt}$ factor but it will not be important in the discussion of this section and we will omit it below).

Now suppose one is close enough to the horizon so that  eq.(\ref{rangea}) is met, but not too close, so that 
\be
\label{ntc}
|\hat{w} \hat {z}| \simeq |\hat{\omega} {\ln(\hat{r}-1) \over B}| \ll 1.
\ee
Then the exponential in eq.(\ref{solse}) can be expanded and the solution in this region becomes  
\be
\label{exppsi}
\psi \simeq 1-i\hat{w}\hat{z}.
\ee
The condition eq.(\ref{ntc}) is 
\be
\label{rangeb}
\hat{r}-1\gg e^{-{B\over \hat{\omega}}}
\ee
which is compatible  with eq.(\ref{rangea}) for $\hat{\omega} \ll 1$. 

Next   consider the region 
\be
\label{condrangec}
1\gg \hat{r}-1 \gg {\hat{w}^2\over A}. 
\ee
In this region  the frequency term in the Schr\"odinger equation is  now subdominant compared to the potential term. 
Moving  even further away from the horizon the frequency will continue to be unimportant   all the way  to the region 
 $\mu \gg \hat{r} \gg 1$ where the coefficient 
$a_1$ is defined. So it is enough to understand the solution in the region eq.(\ref{condrangec}) for establishing that the coefficient $a_1$ is 
unsuppressed by further powers of $\omega$.

By carrying out a  change of variables
\be
\label{chngvar}
x\equiv e^{B \hat{z}\over 2}\sqrt{4A\over B^2} = \sqrt{(\hat{r}-1) 4A\over B^2},
\ee
where in obtaining the last equality we have used the relation eq.(\ref{varza}), 
we can recast the Schr\"odinger equation in the region eq.(\ref{condrangec}) in   the form
\be
\label{recastse}
-(x^2{d^2\psi \over dx^2}+x {d\psi\over dx}) + x^2 \psi=0. 
\ee
This is closely related to the standard  Bessel equation. From eq.(\ref{chngvar}) and eq.(\ref{condrangec}) we see that in this region 
\be
\label{rangecc}
x\ll 1.
\ee
The solution to eq.(\ref{recastse}) then takes the form, 
\be
\label{seform}
\psi=  C_0+C_1 \ln(x)  = C_0+\tilde{C}_1 \hat{z}.
\ee

Now notice that eq.(\ref{exppsi}) and eq.(\ref{seform}) are of the same form. 
There is in fact a good reason for this. As we will see below we can extend the solution  from the region eq.(\ref{rangeb})  
where eq.(\ref{exppsi}) is valid to the region  
eq.(\ref{condrangec})
where eq.(\ref{seform}) is valid by neglecting both the potential and the frequency dependent terms in the Schrodinger equation. 
Neglecting these terms gives a free Schr\"odinger equation at zero energy, 
\be
\label{extrase}
{d^2\psi \over d\hat{z}^2}=0, 
\ee
with the solution
which agrees with eq.(\ref{exppsi}), eq.(\ref{seform}).  

The coefficients $C_0$ and ${\tilde C}_1$ can  therefore be fixed by equating    eq.(\ref{exppsi}) and eq.(\ref{seform}) giving
\be
\label{valca}
C_0=1, \\ \tilde{C}_1= -i \hat{\omega}
\ee
In the region eq.(\ref{condrangec}) it follows from eq.(\ref{varza}) that 
\be
\label{rata}
|\tilde{C_1} z| \sim |\hat{\omega} \ln(\hat{r}-1)| \le |\hat{\omega} \ln(\hat{\omega})| \ll 1,
\ee
where the last inequality follows from the fact that $\hat{\omega} \ll 1$. 
Thus to good approximation we can take 
\be
\label{gapp}
\psi=C_0=1
\ee
in this region. 

We see therefore that the solution is of order unity in this region (without any power law suppression by a  factor of  $\hat{\omega})$.
 And it follows then that 
 going further away from the horizon to the region where $\mu/T  \gg \hat{r} \gg 1$ the coefficient  $a_1$ will also be of order unity. 

To complete the argument let us discuss how to extend the solution from the region eq.(\ref{rangeb}) to eq.(\ref{condrangec}). 
Choose a point with coordinate
\be
\label{p1}
\hat{r}_1-1=c_1 {\hat{\omega}^2\over A}. 
\ee
Here   $c_1$ is a constant which does not scale with $\hat{\omega}$ and  
 meets the condition $c_1 \ll 1$ so that the condition eq.(\ref{rangea}) is met. Since $\hat{\omega} \ll 1$ and $c_1$ does not scale 
with $\hat{\omega}$ we see that   eq.(\ref{rangeb})  is also  met 
and this point lies in the region eq.(\ref{rangeb}). 
Next choose a second point  with coordinate
\be
\label{p1b}
\hat{r}_2-1=c_2 {\hat{\omega}^2\over A}, c_2 \gg 1
\ee
such that $\hat{r}_2 \ll 1$. This point lies in the region eq.(\ref{condrangec}). 
Using eq.(\ref{varza}) we see that the change in $\hat{z}$ in going from $\hat{r}_1$ to $\hat{r}_2$ is 
\be
\label{delz}
\delta{\hat{z}}={1\over B}\ln({c_2\over c_1})
\ee
and is independent of $\hat{\omega}$.

For the frequency dependent term in the Schrodinger equation to be neglected in the process of continuing the solution from $\hat{r}_1$ to
$\hat{r}_2$, the condition  
\be
\label{condnea}
\omega^2 (\delta\hat{z})^2  \ll 1
\ee
must be met. Since $\hat{\omega}\ll 1$ we see that this is true. 
Similarly for the potential dependent term to be negligible the condition 
\be
\label{potcond sea}
V(z) (\delta\hat{z})^2 \sim (\hat{r}-1) (\delta\hat{z})^2 \sim \hat{\omega}^2 (\delta\hat{z})^2 \ll 1
\ee
must be met. This condition is also true, thereby completing the argument.

\section{Appendix B}
Here we   discuss how eq.(\ref{tttwo}) is obtained.
In AdS/CFT the metric is dual to the boundary stress tensor. So eq.(\ref{tttwo}) is obtained by doing a bulk path integral with a fixed boundary
metric and then obtaining the two-point function from it. It is well known that after using  the  equations of motion,
the resulting answer is obtained in terms of the extrinsic curvature of the boundary. In the $SL(2,R)$ invariant case we are dealing with here,
this calculation is particularly simple  since the metric is invariant under $SL(2,R)$. Thus one can work in the purely electric case which is a considerable simplification. This gives the    result eq.(\ref{tttwo}) as we will see shortly.
Transforming to the dyonic frame then keeps the result unchanged since the energy density is invariant.

To calculate eq.(\ref{tttwo}) in the purely electric case we go back to eq.(\ref{exttx}) but now are  more careful since a non-normalisable mode
for $g_{tx}$ is also turned on.
This requires the first subleading corrections in $a^2,b^2$ to be kept,
\begin{eqnarray}
\label{fcorrmet}
a^2 & = & r^2(1-{\kappa^2 \rho\over   r^3}) \\
b^2 & =  & r^2 + \cdots 
\end{eqnarray}
Here we have reinstated the factors of $\kappa^2$; the action eq.(\ref{acsl}) has an overall factor of $2\kappa^2$ in front of it.
We are also working in units where radius of AdS space is set to unity $L=1$. The ellipses on the rhs 
of the equation for $b^2$
 indicate corrections which fall sufficiently  fast and can be   neglected in the  calculation below.  
Keeping these corrections in eq. (\ref{exttx}) leads to 
\begin{eqnarray}
\label{nttx3}
\langle T_{tx} \rangle & = & ({1\over 2\kappa^2}) [ a^3\partial_r({g_{tx}\over a^2})+ 2 g_{tx}(\partial_r a -1)] \\
  && = ({1\over 2\kappa^2}) [ a^3\partial_r({g_{tx}\over a^2}) + 2 {g_{tx} \kappa^2 \rho \over r^3}]
\end{eqnarray}
Eq.(\ref{newvalc}) then  becomes
\be
\label{altnewvalc}
\partial_r({g_{tx}\over a^2}){a^2\over b^2}+{g_{tx}\over a^2}({a^2\over b^2})'={2\over i \omega}({a^2\over b^2})[2\lambda_2-F_{rt} F_{tx} g^{tt}]
\ee
Leading to
\be
\label{twopttta}
\langle T_{tx} \rangle=-{\rho g_{tx} \over 2   r^3 }  - {4\over i \omega}[{Q_e' E_x'\over a}]
\ee
Now differentiating with respect to $g_{tx}$ and converting to gauge theory variables gives eq.(\ref{tttwo}) for $i=j=x$.
 In the absence of a magnetic field there is no  cross-talk  between  the $g_{xt}$ and $g_{yt}$  perturbations so 
$\langle T_{tx}, T_{ty} \rangle =0$, which is the second relation contained in eq.(\ref{tttwo}). 

\section{Appendix C}
Here we provide a more detailed analysis of some of the results discussed in  \S8.
 
For the case where $4K+2+2\beta>0$ and $P>2$, which was discussed in \S8.2, the equation for the axion eq.(\ref{sttwo})
 has two solutions in the near-horizon region where  $x\rightarrow \infty$. Both solutions  can be expressed as a power series in $x$. 
The first is
\be
\label{fsolac}
\lambda_1=c_1 x+c_2x^{\alpha} + \cdots. 
\ee
For the second term on rhs to be subdominant compared to the first  when $x\rightarrow \infty$ 
\be
\label{condapc}
\alpha<1.
\ee
Substituting  eq.(\ref{fsolac}) in eq.(\ref{sttwo}) and equating powers of $x$ gives, 
\be
\label{px}
\alpha=3-P. 
\ee
Requiring that condition eq.(\ref{condapc}) is met gives, 
\be
\label{condapc2}
P>2
\ee
which is indeed true. 
This solution blows up as $x\rightarrow \infty$. 

The second solution to eq.(\ref{sttwo}) is
\be
\label{secsol}
\lambda_1=c_0+c_1x^\alpha + \cdots,
\ee
with the condition,  
\be
\label{condsecsola}
\alpha<0.
\ee
Substituting in eq.(\ref{sttwo}) and equating powers of $x$   gives  
\be
\label{valalphaappc}
\alpha=2-P,
\ee
so that eq.(\ref{condsecsola}) is again met. 
Equating coefficients   determines $c_1$ in terms of  $c_0$.
 
In summary we learn that for the
axion to be non-zero (i.e. away from its attractor value ) and for it to not  blow up at the horizon,  it must be of the form 
eq.(\ref{secsol}) with $c_0$ non-vanishing.  Thus, $\lambda_1$ does not vanish as $x\rightarrow \infty$ and we do not get attractor 
behaviour in this case. 

Next consider the case where $4K+2+2\beta<0$ and $P<2$, also discussed in \S8.2.
Now $x\rightarrow 0$ at the horizon. 
An analysis, very similar to the one above,  shows that there is a solution to the axion equation eq.(\ref{sttwo})
  of the form 
\be
\label{solaxcasec}
\lambda_1=c_1 x + c_2 x^\alpha + \cdots. 
\ee
with $\alpha>1$.
In this solution $\lambda_1\rightarrow 0$ and does indeed reach its attractor value at the horizon. 
However, as was discussed in \S7.2 one must examine the backreaction due to the varying  axion  on the
other  equations of motion.   

One of the Einstein equations is
\be
\label{oneee}
{\partial_r^2 b \over b}=-(\partial_r\phi)^2-{1\over 4}e^{4\phi} (\partial_r \lambda_1)^2
\ee
In the purely magnetic dilaton black brane the axion contribution  vanishes and the lhs is balanced by the 
first term on the rhs with both terms scaling like $1/r^2$. From the solution for the axion eq.(\ref{solaxcasec})
and eq.(\ref{dilpmag}) and eq.(\ref{defax})
 we see that the second term on the rhs of eq.(\ref{oneee}) scales like  ${1\over r^2 r^{4K+4\beta+2}}$.
It is   easy to see using eq.(\ref{constants}) that $4K+4\beta+2>0$ and therefore the axion contribution always dominates
for small enough $r$. This establishes  that the axion  backreaction cannot be neglected.

\section{Appendix D}

In this appendix, we study  the theory described by the action (\ref{generalaction}) and attempt to find purely power-law 
near-horizon solutions for $\alpha \neq -1$.


The equations of motion for the system can be
obtained following section 6 of \cite{attractors}. Using our usual metric ansatz:
\begin{equation}
\label{ansatzb}
ds^2 = -a_R(r)^2 ~dt^2 + a_R(r)^{-2} ~dr^2 + b(r)^2 ~(dx^2 + dy^2),
\end{equation}
we obtain a one-dimensional action,
\bea
S= \int dr
  \left(2-(a_R^{2}b^{2})^{''}-2a_R^{2}bb^{''}-2a_R^{2}b^{2}(\partial_{r}\phi)^{2}
  -\frac{1}{2}e^{4\phi}a_R^2b^2(\partial_r a)^2-2\frac{V_{eff}}{b^{2}}
    +\frac{3b^{2}}{L^{2}}\right)  \quad \quad
\eea

The equations of motion arising from this action are:
\bea
2 \partial_r \left( a_R^2 b^2 \partial_r \phi \right) &=& \frac{\partial_\phi V_{eff}(\phi, a)}{ b^2} + e^{4 \phi} a_R^2 b^2 \left( \partial_r a \right)^2 \,, \\
\label{originalaxioneq}
\partial_r \left(e^{4 \phi} a_R^2 b^2 \partial_r a \right) &=& 2 \,\frac{\partial_{a} V_{eff}(\phi, a)}{ b^2} \,,\\
\frac{\partial_r^2 b}{b} &=& - (\partial_r \phi)^2  - \frac{1}{4} e^{4 \phi}(\partial a)^2 \,, \\
\partial_r^2 \left( a_R^2 b^2 \right) &=& -4 \Lambda b^2 \,.
\eea

We look for solutions of the following form:
\bea
&& ds^2 = - C_R^2 \left(r-r_h \right)^2 dt^2 +  C_R^{-2} \left(r-r_h \right)^{-2} dr^2 +  C_{\beta}^2 \left(r-r_h \right)^{2 \beta} \left( dx^2 +
dy^2 \right) \,, \nonumber \\
&& e^{\phi} = C_{\phi} \left(r-r_h \right)^{K}  \,, \quad   \lambda_1 = a = C_a  \left(r-r_h \right)^{\gamma} + a_* \
\label{defofKgamma}
\eea
where $C_R, C_{\beta}, C_{\phi}, C_a$ are constants, and $a_*\equiv\lambda_{1*}$ is the attractor value for axion given by
(\ref{attgen}), which minimises the effective potential
\bea
\label{effectivepotforappendix}
V_{eff}(\phi, a) = e^{- 2 \alpha \phi} \left(Q_e - a Q_m \right)^2 + e^{2 \alpha \phi} Q_m^2 \,.
\eea

Substituting this ansatz into the above equations of motion yields, 
\bea
\label{dilatoneq}
2 (1 + 2 \beta)C_R^2 C_\beta^2 K \left( r-r_h \right)^{2 \beta} = - 2 \alpha C_\beta^{-2} C_\phi^{-2 \alpha} C_a^2 Q_m^2 \left( r-r_h \right)^{-2 \alpha K + 2 \gamma - 2 \beta} \quad \quad \quad \quad  \nonumber \\
  +2  \alpha C_\beta^{-2}  C_\phi^{2 \alpha} \left( r-r_h \right)^{2 \alpha K - 2 \beta}  +  C_\phi^4 C_R^2 C_\beta^2 C_a^2 \gamma^2
 \left( r-r_h \right)^{4 K + 2 \beta + 2 \gamma}  \,,  \quad \quad  \quad \\
\label{axioneq}
\gamma (4 K + 2 \beta + \gamma + 1) C_\phi^4 C_R^2 C_\beta^2  \left( r-r_h \right)^{4 K + 4 \beta} = 4 \, Q_m^2 C_\beta^{-2} C_\phi^{-2 \alpha} \left( r-r_h \right)^{- 2 \alpha K } \,, \quad \quad \quad \\
\label{metriceq}
\beta (\beta - 1) \left( r-r_h \right)^{-2}  = - K^2 \left( r-r_h \right)^{-2} - \frac{\gamma^2 C_{\phi}^4 C_{a}^2}{4} \left( r-r_h \right)^{4 K + 2 \gamma - 2}  \,,\quad \quad \quad   \,\, \\
\label{trivialeq}
(2\beta+2) (2\beta + 1) C_R^2\left( r-r_h \right)^{2\beta}  = -4 \Lambda \left( r-r_h \right)^{2\beta}  \,.\quad \quad \, \quad \, \quad \, \quad \, \quad
\eea

We now seek to determine if a solution for the coefficients $C_R, C_\phi, C_a, C_\beta$ and  exponents $\beta, \gamma, K$ exists that solves the above equations for general values of $\alpha \neq 0, -1$.

For the attractor mechanism to work, the dilaton and axion must flow to minima of the effective potential. In terms of our power law ansatz,
\bea
\partial_a V_{eff} &=& - Q_m e^{- 2 \alpha \phi} (Q_e - a Q_m)  = Q_m^2 C_{\phi}^{-2 \alpha} C_a  \left(r-r_h \right)^{- 2 \alpha K + \gamma}   \,,\\
\partial_\phi V_{eff} &=& - 2 \alpha C_{\phi}^{-2 \alpha} C_a^2 \left(r-r_h \right)^{- 2 \alpha K + 2 \gamma} + Q_m^2 C_\phi^{2 \alpha}\left(r-r_h \right)^{2 \alpha K}  \,.
\eea
Requiring $\partial_a V_{eff} \to 0$ and $\partial_\phi V_{eff} \to 0$  as $r \to r_H$ then results in the inequality
\bea
\label{gammaequality}
\gamma > 2 \alpha K
> 0 \,.
\eea

We first consider the metric equation (\ref{metriceq}). To satisfy this equation in the near-horizon region where $r \to r_h$, we must impose the following inequality:
\bea
\label{uneqlity}
4 K + 2 \gamma \ge 0
\eea 

We next consider the axion equation of motion (\ref{axioneq}). Note that if the coefficient on the LHS vanishes, i.e., $4 K + 2 \beta + \gamma + 1 = 0$, we must also take into account the subleading behaviour for the axion. We, therefore, separately consider case {\bf A)} $4 K + 2 \beta + \gamma + 1 \neq 0$ and case
{\bf B)} $4 K + 2 \beta + \gamma + 1 = 0$. (Assuming $\beta \neq -1/2, -1, 0, 1$, the inequality (\ref{gammaequality}) implies all other coefficients are nonzero.)

{\bf A)} Case: $4 K + 2 \beta + \gamma + 1 \neq 0$.

(\ref{axioneq}) gives the relation
\bea
 4 K + 4 \beta =  - 2 \alpha K  \,.
\label{axiontwo}
\eea
We must separately consider the case where (\ref{uneqlity}) is a strict inequality, and the special case where $4 K + 2 \gamma =0$.

{\bf A-1)} Case: $4 K + 2 \beta + \gamma + 1 \neq 0$ with $4 K + 2 \gamma > 0$.

We consider the dilaton equation of motion (\ref{dilatoneq}). From eq. (\ref{axiontwo}) and the inequality (\ref{uneqlity}), the LHS of (\ref{dilatoneq}) always dominates the first term and third terms on the RHS of (\ref{dilatoneq}) in the $r -r_h \to 0 $ limit. Therefore, LHS must be balanced with
the second term on the RHS, which gives
\bea
\label{anotherconstraint}
4 \beta = 2 \alpha K
\eea
For $\alpha \neq -1$,  (\ref{axiontwo}) and (\ref{anotherconstraint}) admit no solution. 

Of course, for $\alpha = -1$ the power-law solution exists. This is consistent with the results of \S6, \S7 --
for $\alpha = -1$, the Lagrangian has an SL(2,R) duality symmetry, so  starting from the purely electric (or purely magnetic) case with no axion,
we can construct dyonic black brane solution where the flow of the axion is governed by a power-law using SL(2,R) duality transformations.

{\bf A-2)} Case: $4 K + 2 \beta + \gamma + 1 \neq 0$ with $4 K + 2 \gamma = 0$.

In this case,
\bea
\label{gammaKrela}
\gamma = - 2 K > 0 \,
\eea
and all terms in the dilaton equation (\ref{dilatoneq}) have exponent equal to $2\beta$ except the second term on the right hand side, whose exponent is $2\alpha K - 2 \beta$. If this second term were to dominate the others, we would not be able to satisfy the dilaton equation (unless $\alpha = 0$). Therefore, we require that the exponent of the second term is greater than or equal to $2\beta$, {\it i.e.,}
\bea
4 \beta \le 2 \alpha K \,.
\eea
Using (\ref{axiontwo}) with $K< 0$ and $\alpha \neq -1$, this implies
\bea
\alpha < -1 \,.
\eea
On the other hand, (\ref{gammaequality}) and (\ref{gammaKrela}) give
\bea
\alpha > -1 \,,
\eea
which is a contradiction. Again, we do not find a power-law solution.

{\bf B)} Case: $4 K + 2 \beta + \gamma + 1 = 0$.

In this case, we have to consider the sub-leading correction in the axion equation (\ref{originalaxioneq}) instead of (\ref{axioneq}). Again we consider separately the
cases when (\ref{uneqlity}) is an equality or a strict inequality.

{\bf B-1)} Case: $4 K + 2 \beta + \gamma + 1 = 0$  with $4 K + 2 \gamma > 0$.

By (\ref{gammaequality}), the second term on the RHS of the dilaton equation (\ref{dilatoneq}) is always greater than the first term, and since $4 K + 2 \gamma > 0$, the LHS is always greater than the third term on the RHS of (\ref{dilatoneq}). Therefore, the term on the LHS must be balanced with the second term in the RHS, yielding
\bea
\label{Keqone}
2 \beta =  \alpha K \,.
\eea

Using $4 K + 2 \gamma > 0$, matching the coefficients of the metric equation (\ref{metriceq}) gives
\bea
\label{Keqtwo}
\beta (\beta - 1) = - K^2
\eea

Using these two results (\ref{Keqone}) and (\ref{Keqtwo}) and $4 K + 2 \beta + \gamma + 1 = 0$, we can solve for nonzero $K$ and $\beta$ in terms of $\alpha$:
\bea
K = \frac{2 \alpha}{\alpha^2 + 4}  \,,\quad \beta = \frac{\alpha^2}{\alpha^2 + 4}  \,, \quad
\gamma = -3 + 8 \frac{1 - \alpha}{\alpha^2 + 4}
\eea

However, for any real $\alpha$,
\bea
4 K + 2 \gamma =  - \frac{ 6 (\alpha + 2/3)^2 + 16/3}{\alpha^2 + 4} < 0 \,,
\eea
which is in contradiction with the assumption that $4 K + 2 \gamma > 0$. Therefore, there is no power-law solution in this case.

{\bf B-2)} Case: $4 K + 2 \beta + \gamma + 1 = 0$  with $4 K + 2 \gamma = 0$.

%
%
%

These two conditions imply that
\bea
\beta = -\frac{1 + 2 K}{2}  \,, \quad \gamma = - 2 K
\eea

We now attempt to solve for the coefficients in the metric equation (\ref{metriceq}). We obtain:
\bea
\gamma^2 C_a^2 C_\phi^4/4 = -\beta (\beta - 1) - K^2 = -2 ( K + 1/2)^2 - 1/4 < 0
\eea
Therefore, (\ref{metriceq})
cannot be satisfied and there is no power law solution.

The above analysis seems to be fairly exhaustive, and we conclude that no power law solution exists for $\alpha \neq -1$. 
The special cases of $\beta= -1/2, -1, 0, 1$ were not studied -- it would be interesting to see if one of these special cases permits a power law solution for some other values $\alpha$. 
It is also conceivable that for other values of $\alpha$, 
 the axion approaches the attractor value as a power law in $r-r_h$ with additional logarithmic corrections. 
We do not, however, explore such possibilities here.




\end{document}